\documentclass[english]{article}
\usepackage[T1]{fontenc}
\usepackage[utf8]{luainputenc}
\usepackage{geometry}
\usepackage{babel}
\usepackage{float}
\usepackage{textcomp}
\usepackage{amsmath}
\usepackage{amsthm}
\usepackage{graphicx}
\usepackage{setspace}
\usepackage[unicode=true,pdfusetitle,
 bookmarks=true,bookmarksnumbered=false,bookmarksopen=false,
 breaklinks=false,pdfborder={0 0 1},backref=false,colorlinks=false]
 {hyperref}

\makeatletter
\theoremstyle{plain}
\newtheorem{lyxalgorithm}{\protect\algorithmname}

\makeatother

\providecommand{\algorithmname}{Algorithm}

\begin{document}
\begin{center}
\textbf{\LARGE{}To Trade Or Not To Trade: Cascading Waterfall Round
Robin Rebalancing Mechanism for Cryptocurrencies}{\LARGE\par}
\par\end{center}

\begin{center}
\textbf{\large{}Ravi Kashyap (ravi.kashyap@stern.nyu.edu)}\footnote{\begin{doublespace}
Numerous seminar participants, particularly at a few meetings of the
econometric society and various finance organizations, provided suggestions
to improve the paper. The following individuals have been a constant
source of inputs and encouragement: Dr. Yong Wang, Dr. Isabel Yan,
Dr. Vikas Kakkar, Dr. Fred Kwan, Dr. Costel Daniel Andonie, Dr. Guangwu
Liu, Dr. Jeff Hong, Dr. Humphrey Tung and Dr. Xu Han at the City University
of Hong Kong. The views and opinions expressed in this article, along
with any mistakes, are mine alone and do not necessarily reflect the
official policy or position of either of my affiliations or any other
agency.
\end{doublespace}
}
\par\end{center}

\begin{center}
\textbf{\large{}Estonian Business School / City University of Hong
Kong}{\large\par}
\par\end{center}

\begin{center}
\today
\par\end{center}

\begin{center}
\par\end{center}

\begin{center}
Journal of Economic Literature Codes: G11 Investment Decisions; D81
Criteria for Decision-Making under Risk and Uncertainty; C32 Time-Series
Models; B23 Econometrics, Quantitative and Mathematical Studies; D8:
Information, Knowledge, and Uncertainty; O3 Innovation • Research
and Development • Technological Change • Intellectual Property Rights;
\par\end{center}

\begin{center}
under uncertainty in the context of artificial intelligence
\par\end{center}

\begin{doublespace}
\begin{center}
\pagebreak{}
\par\end{center}
\end{doublespace}

\begin{center}
\tableofcontents{}\pagebreak{}
\par\end{center}

\begin{doublespace}
\begin{center}
\listoffigures 
\par\end{center}

\begin{center}
\listoftables 
\par\end{center}
\end{doublespace}

\begin{singlespace}
\begin{center}
\pagebreak{}
\par\end{center}
\end{singlespace}
\begin{doublespace}

\section{Abstract}
\end{doublespace}

We have designed an innovative portfolio rebalancing mechanism termed
the Cascading Waterfall Round Robin Mechanism. This algorithmic approach
recommends an ideal size and number of trades for each asset during
the periodic rebalancing process, factoring in the gas fee and slippage.
The essence of the model we have created gives indications regarding
whether trades should be made on individual assets depending on the
uncertainty in the micro - asset level characteristics - and macro
- aggregate market factors - environments. In the hyper-volatile crypto
market, our approach to daily rebalancing will benefit from volatility.
Price movements will cause our algorithm to buy assets that drop in
prices and sell as they soar. In fact, the buying and selling happen
only when certain boundaries are crossed in order to weed out any
market noise and ensure sound trade execution. We have provided several
numerical examples to illustrate the steps - including the calculation
of several intermediate variables - of our rebalancing mechanism.
The Algorithm we have developed can be easily applied outside blockchain
to investment funds across all asset classes at any trading frequency
and rebalancing duration.

\subsection{Shakespeare As A Crypto Trader}

\noindent \textbf{\textit{To Trade Or Not To Trade, that is the Question,}}

\noindent \textbf{\textit{Whether an Optimizer can Yield the Answer,}}

\noindent \textbf{\textit{Against the Spikes and Crashes of Markets
Gone Wild,}}

\noindent \textbf{\textit{To Quench One’s Thirst before Liquidity
Runs Dry,}}

\noindent \textbf{\textit{Or Wait till the Tide of Momentum turns
Mild.}}

This is inspired by Prince Hamlet's soliloquy in the works of Shakespeare:
\textquotedbl To be or not to be; that is the question\textquotedbl{}
(End-note \ref{EN:To-Be-Not-To-Be}; Bradley 1991).

\section{\label{sec:Introduction:-Costs,-Constraints}Introduction: Costs,
Constraints and Countless Critical Considerations for Portfolio Rebalancing}

Rebalancing is a method of ensuring that a financial portfolio - comprised
of investment assets - stays aligned with its intended risk and return
objectives (Ross et al., 1999; Tokat \& Wicas 2007; Elton et al.,
2009; End-note \ref{enu:In-finance-Rebalancing}). A portfolio is
created based on allocations of wealth to various assets to achieve
certain economic goals. Deviations from expectations - in terms of
risk and return, as reflected by movements in the asset prices and
changes to portfolio weights - happen due to changes in market conditions
and / or asset characteristics and / or due to changes in the intentions
themselves. 

The essence of rebalancing simply involves decreasing the exposure
to assets that have become a bigger proportion of the portfolio and
increasing the exposure to assets that have smaller allocations to
them - as compared to their respective holdings at a previous point
in time. Clearly the funds obtained from selling some securities can
be used for the purchase of other instruments. In addition additional
funds can be used to buy assets - to enhance the portfolio and change
the asset weights with the passage of time, without selling any assets
- or profits can be taken out - to trim the portfolio - by selling
some assets and not making any new asset purchases.

The necessity to rebalance asset holdings is widely acknowledged -
and documented - for both risk management in investment portfolios
and for capital structure decisions - which are also risk driven -
in corporate finance circles (Leary \& Roberts 2005; Cook \& Tang
2010; Rastad 2016; Chauhan \& Huseynov 2018; Juelsrud \& Wold 2020;
An et al., 2021). There are numerous mutually educational lessons
from both these approaches - investment management and capital structure
decisions - for either side that seeks to change their asset compositions.
In this paper, we are focusing on the risk management of investment
vehicles holding several assets - which tend to happen more frequently
compared to capital structure rebalancing. 

The proponents of rebalancing advocate its use due to efficacies in
improving returns and reducing risks. Maeso \& Martellini (2020) find
that - after controlling for various factor risk exposures (Cochrane
2009; End-note \ref{enu:In-finance,-risk-Factors}) - the average
outperformance of the rebalanced portfolio when compared to the corresponding
buy-and-hold portfolio remains substantial at an annualized level
- above 100 basis points over a 5-year time horizon for stocks in
the S\&P 500 universe. 

Numerous studies identify variables that can influence the decision
to rebalance and provide conceptual frameworks - including several
practical suggestions - for developing rebalancing strategies. Such
techniques are based on gauging changes in the financial landscape
and asset properties while being cognizant of the costs of rebalancing,
the risk tolerance and investment time horizon of the investors. Many
sophisticated methodologies - involving complex resource intensive
calculations and operational efforts - have been created in contrast
to the more commonly used simpler alternatives - which tend to be
highly generic limiting their effectiveness for specific investment
goals (Masters 2003; Sun et al., 2006; Tokat \& Wicas 2007; Stutzer
2010). 

Due to this possibility of expending varying levels of efforts for
rebalancing, we can expect differences in rebalancing approaches to
be more prevalent between more sophisticated institutional investors
and less informed individual investors. Calvet et al., (2009) investigate
the dynamics of individual portfolios using household data from Sweden.
They find that that households primarily rebalance by increasing their
sales of risks assets - and reducing their asset buys - when they
have higher than average returns, and by increasing their purchases
of risky assets when they have lower than average returns. 

Each asset in a portfolio contributes to the overall compound return
of the portfolio - termed the return contribution of that asset -
with the rebalancing mandate of keeping a constant weight across each
asset. This share of the asset towards the portfolio compound return
- the return contribution - exceeds the asset's compound return by
an additional amount which is know as the diversification return.
The diversification return of the overall portfolio is the weighted
average of individual assets’ diversification returns. Measuring diversification
return can be a useful way to determine the advantages of rebalancing
- despite several misunderstandings in assessing the benefits and
the need to pay attention to the nuances of the situations where rebalancing
actually yields benefits (Willenbrock 2011; Bouchey et al., 2012;
Qian 2012; Chambers \& Zdanowicz 2014; Hallerbach 2014). 

Several tools and techniques have been developed which facilitate
rebalancing specific to an investment horizon while taking into account
the corresponding transaction costs that are generated (Woodside-Oriakh
et al., 2013; Cuthbertson et al., 2016; ). More realistic multi-period
analysis such as in Guastaroba et al., (2009) study the effect of
fixed and proportional transaction costs on rebalancing in a multi-period
setting. Kimball et al., (2020) develop an overlapping generations
model of optimal rebalancing where agents differ in age and risk tolerance. 

Yu \& Lee (2011) compare several portfolio rebalancing models based
on different combinations of criteria such as transaction cost, risk,
return, short selling, skewness, and kurtosis. Rebalancing can be
more strategic - by not following set time schedules - based on trends
in the movement of asset prices, and in particular delaying rebalancing
when markets are moving downwards can reduce the negative impact of
drawdowns (Rattray et al., 2020). Another aspect is the extent of
international securities in a portfolio versus the domestic holdings
- and hence the influence of foreign exchange and returns from foreign
securities - which influence rebalancing and the corresponding capital
flows (Stein et al., 2009; Camanho et al., 2017; 2022).

\subsection{\label{subsec:Blockchain-Based-Rebalancing}Blockchain Based Rebalancing
and Trade Execution}

Kashyap (2022) describes several innovations geared at bringing better
wealth appreciation and risk management methodologies to the decentralized
investment landscape. These innovations create novel investment techniques
- in addition to modifying many traditional finance principles where
necessary - aimed at delivering superior risk adjusted returns to
everyone - accessible to the masses due to the use of decentralized
technology. 

In this article, we take a closer look at the  trade execution innovations
we have brought to the decentralized finance (DeFi - Zetzsche et al.,
2020; Jensen et al., 2021; End-note \ref{enu:Decentralized-finance})
space to be able to securely and efficiently execute trades to rebalance
portfolios on a daily basis or even at an intraday frequency. The
Algorithm we detail below can be easily applied outside blockchain
to investment funds across all asset classes at any trading frequency
and rebalancing duration.

Our rebalancing algorithm can be summarized in a few words as the
“Cascading Waterfall Round Robin Mechanism”. To describe how this
algorithm works, we first start by assigning to each asset in our
portfolio a certain capacity to hold funds. This capacity is the result
of several calculations that depend upon: 1) the risk and return properties
of each asset, 2) how the asset prices vary in comparison to other
assets in the portfolio, and 3) the amount of funds collected for
investment or the total requests for redemption. 

In Section (\ref{sec:Weight-CE}) we will go into greater detail regarding
the use of risk and return characteristics to arrive at the capacity
for each asset. Once the capacity is determined, we check how much
of that capacity is utilized. This gives us an idea of how much money
we can put into each individual asset when we have to invest money
across our assets. Likewise, it also tells us how much we can pull
out of each asset if we need to withdraw money from assets. Next,
we distribute funds among the assets - or redeem funds from the assets
- in a circular manner - or round robin fashion - till the full capacity
of each asset is reached. As the capacity on one asset reaches its
full limit, the funds start trickling down to the next asset, similar
to a waterfall. The reverse happens when redemptions are to be fulfilled.
Hence the name, “Cascading Waterfall Round Robin Mechanism”.

After the trade execution schedule is decided, we need to consider
the transaction costs of completing the buy and sell orders. There
are two main implicit costs we face at this stage. First are the gas
fees for each transaction we execute (Pierro et al., 2020; 2022).
The second cost of doing trades is known as slippage or market impact
(Bertsimas \& Lo 1998; Almgren \& Chriss 2001; Karastilo 2020). The
gas fees depend on a number of factors, such as the time of execution
and the network on which we are transacting (Zarir, et. al. 2021;
Donmez \& Karaivanov 2022). The slippage will depend on the size of
our trades in comparison to the sources of liquidity chosen for doing
the trades. If we have more trades, the total gas costs will increase.
If we have larger trade sizes, the slippage will increase.

The quintessential trading conundrum in traditional finance is whether
(and how much) to trade during a given interval or wait for the next
interval when the price momentum is more favorable to the direction
of trading. The problem is compounded in the crypto domain since we
have to factor in the gas fees. The optimal amount of funds, to be
moved in and out of assets, is determined by the dual objectives of
minimizing both gas fees and slippage or market impact costs. 

We perform asset level calculations which are coupled with our “Cascading
Waterfall Round Robin Mechanism” to arrive at recommended minimum
and maximum trade sizes (Section \ref{subsec:Trading-Block-Size}).
These trade size recommendations ensure that our fund managers can
adhere to certain security guidelines, when funds need to be moved
into and out of assets according to a blockchain specific security
blue print described in Katarina (2023). The goal of strengthening
security is achieved without creating bottlenecks for trading since
fund movements correspond to trade size restrictions.

The calculation of asset capacities and the rebalancing methodology
are among the most central elements of any investment process. It
is no different in the overall plan we have created to bring better
risk management to decentralized finance (Kashyap 2022). If anything
they - asset capacity calculations and the rebalancing methodology
- become more important given that our overall process adheres to
strict risk metrics. This emphasis on rigorous risk management necessitated
that we had to build several new techniques geared towards overcoming
the additional challenges in the decentralized space. The first piece
of work we undertook was related to rebalancing underscoring the importance
of this component towards accomplishing superior risk management on
blockchain.

\subsection{\label{subsec:Outline-of-the}Outline of the Sections Arranged Inline}

Section (\ref{sec:Introduction:-Costs,-Constraints}) which we have
already seen, provides an introductory overview of rebalancing in
investment funds and an intuitive description of the mechanism we
have developed for blockchain investment vehicles. Section (\ref{sec:The-Cascading-Waterfall-Rebalancing})
gives a step by step algorithm that performs rebalancing after factoring
in various constraints related to trade execution in a decentralized
environment. Section (\ref{sec:Rebalancing-Flow-Chart}) has the flow
charts related to the material discussed in Section (\ref{sec:The-Cascading-Waterfall-Rebalancing}).
The diagram in Section (\ref{sec:Rebalancing-Flow-Chart}) is given
for completion and for helping readers obtain a better understanding
of the concepts involved. 

Section (\ref{sec:Gas-Fees-versus-Slippage}) is a discussion of some
suggested methods to arrive at minimum and maximum trade execution
sizes on any asset. Section (\ref{sec:Weight-CE}) gives a summary
of weight calculation techniques that can help determine the minimum
and maximum investment capacity across each asset. Sections (\ref{sec:Gas-Fees-versus-Slippage};
\ref{sec:Weight-CE}) provide inputs to the calculations being done
in the main rebalancing algorithm given in Section (\ref{sec:The-Cascading-Waterfall-Rebalancing}).

Section (\ref{sec:Numerical-Results}) explains the numerical results
we have obtained, which illustrate examples of how our rebalancing
algorithm works and also how our innovations compare to existing rebalancing
techniques. Sections (\ref{sec:Areas-for-Further}; \ref{sec:Conclusion})
suggest further avenues for improvement and the conclusions respectively. 

\section{\label{sec:The-Cascading-Waterfall-Rebalancing}The Cascading Waterfall
Round Robin Rebalancing Algorithm}
\begin{itemize}
\item The following algorithm aims to capture the key concepts of our rebalancing
methodology. This algorithm is one of the earliest pieces any investment
firm should focus on implementing in its entirety. This is because
of the importance of generating an optimal set of trades while minimizing
transaction costs. Hence, it is important to ensure that sufficient
time is allotted - by any fund launching blockchain investment operations
- so that implementation and testing of the corresponding logic can
happen as early as possible. The other components (Sections \ref{sec:Gas-Fees-versus-Slippage};
\ref{sec:Weight-CE}) that interact with the rebalancing piece - to
produce the list of orders - can have simplified implementations initially
and they can be improved over time. 
\item When rebalancing across multiple networks (Ethereum Mainnet, ETH,
or Binance Smart Chain Mainnet, BSC; Cernera et al., 2023; End-notes
\ref{enu:CoinMarketCap}; \ref{enu:Crypto-Ranking}) the following
logic applies only to the assets within each network. For example,
Ethereum Mainnet is likely to be rebalanced only once a day, but BSC
Mainnet will be rebalanced every four hours. This means that every
four hours the amount to deploy or withdraw will be spread across
BSC assets only. The exception is that every 6th rebalancing event,
the assets from Ethereum will also be part of the rebalancing mechanism.
When other networks are added, the same logic applies but perhaps
with different rebalancing intervals. For example, Polygon Network
Mainnet (MATIC), when it is included, can be rebalanced every hour.
The number of hours to rebalance is only an example and it has to
depend on the type of trading strategy. The main point being conveyed
here is that networks with lower gas fees are generally rebalanced
more often.
\item If there are no bridge constraints the below mechanism can be applied
in one go across all assets - spread across various platforms - that
are participating in the rebalancing event. When there are bridge
limitations, Kashyap (2023) provides modifications to the amounts
to be deployed on each of the individual participating networks. It
also has adjustments to provide network specific asset weights. It
is to be understood that the relevant steps in Kashyap (2023) are
to be carried out before the algorithm below is attempted when there
are multiple networks.
\item Kashyap (2023) considers limitations on bridge capacities when multiple
networks are included in a rebalancing event. It provides a mechanism
to modify asset weights to be network specific when the same asset
is available on multiple networks. The weights to be used in this
section for the assets in any network, with bridge constraints, are
the ones calculated in one of the steps in the algorithm given in
Kashyap (2023). The amounts to be deployed on each network, after
considering the bridge capacity constraints, are also calculated.
\item This rebalancing logic can be applied even to the liquidity pools
and indices (Xu \& Feng 2022; End-note \ref{enu:Types-Yield-Enhancement-Services})
- and the strategies they hold - at a corresponding rebalancing frequency.
If the vault or index contains other indices, strategies or complex
components, the principle behind this rebalancing mechanism will work
in a similar manner.
\item What this mechanism does is it will decide how much allocation (or
dollar amount) a particular asset - or other strategy - will get across
the overall portfolio. Once this decision is made. A strategy specific
component will distribute that dollar amount to the various pieces
within that strategy. A recursive approach can also be undertaken
where subcomponents can apply this mechanism further. But it is simpler
to have this rebalancing approach for the top level portfolio and
rely on simpler distribution mechanisms for the component assets and
strategies.
\item Many of the calculation steps below could be combined when iterating
through the list of assets in a loop. Other computational overhead
improvements are also recommended accordingly. The below steps have
been broken down to ensure that the main ideas of the mechanism are
easily understood. 
\item $\left(minw_{it},idealw_{it},maxw_{it}\right)$ are the minimum, ideal
and maximum weights recommended for asset $i$ at time $t$. Note
that, $\left(minw_{it}\leq idealw_{it}\leq maxw_{it}\right)$.
\item $\left(mins_{it},maxs_{it}\right)$ represent the minimum and maximum
order size for asset $i$ at time $t$. Note that,$\left(mins_{it}\leq maxs_{it}\right)$.
\item \textbf{Other notation is explained as it occurs in the below steps.
We will add a dictionary of notation in the Appendix where all the
notation and terminology will be listed with suitable explanations.}
\item Coins, tokens, vaults and any vehicle capable of holding investments
will be considered an asset here. 
\item Note that negative amounts denote outflows (or sells) and positive
amounts denote inflows (buys).
\item There will be a total of $k_{t}$ assets at time $t$. 
\item Rebalancing occurs every $T$ minutes and the following algorithm
will generate an order list to deploy or withdraw money into the holdings
based on the net amount collected since the previous rebalancing event.
Note that rebalancing will occur for different networks at different
time intervals. $N$ is the number of rebalancing events that have
occurred from inception till the present time, $t$. 
\begin{equation}
N=\left\lfloor \frac{t}{T}\right\rfloor 
\end{equation}
\item $\left\lfloor x\right\rfloor $ represents rounding to the nearest
integer less than or equal to $x$. $\left\lceil x\right\rceil $
represents rounding to the nearest integer greater than or equal to
$x$.
\end{itemize}
\begin{lyxalgorithm}
\label{alg:Rebalancing-Algorithm}The following algorithm captures
the summary of the rebalancing mechanism by outlining the following
steps:
\end{lyxalgorithm}
\begin{enumerate}
\item \label{enu:1-Read--for}Read $\left(minw_{it},idealw_{it},maxw_{it}\right)$
for each asset.
\item \label{enu:2-Read--for}Read $\left(mins_{it},maxs_{it}\right)$ for
each asset.
\item \label{enu:3-Get-current-notional}Get current notional amounts invested
across all existing assets, $currentTotalAmount_{t}$. This is done
by adding up the amount in each asset $i$, $currentAmount_{it}$,
using the quantity $q_{it}$ of the asset times its latest price $p_{it}$
at the present time, $t$. Note that, there could be new assets every
rebalancing period, in which case the current notional amounts will
be zero for such assets. \textbf{When an asset is to be sold completely,
its new weight will become zero }$\left(minw_{it}=idealw_{it}=maxw_{it}=0\right)$\textbf{.
When an asset position is to be exited completely, the following min
and max positions (based on the weights) will not apply to such assets,
the asset will be completely liquidated using the minimum and maximum
order size.}
\begin{equation}
currentTotalAmount_{t}=\sum_{i=1}^{k_{t}}\left(q_{it}\right)\left(p_{it}\right)
\end{equation}
\begin{equation}
currentAmount_{it}=\left(q_{it}\right)\left(p_{it}\right)
\end{equation}
\item \label{enu:4-Get-the-net}Get the net new amount to be deployed, $TBDAmount_{t}$
since the last rebalancing event, $N$. If this amount is positive,
then we have a net buy or deposit indicator, $depositIND_{t}$ set
to 1. If this amount is negative, then we have a net sell indicator
or a withdrawal indictor set to 1, $withdrawIND_{t}$. Note that only
one of them is 1 for any rebalance event. That is they are negations
of each other. We can also use only one of them, but some formulae
become simpler and we can eliminate a lot of condition checking if
we have both indicators.
\begin{equation}
depositIND_{t}=\begin{cases}
1, & TBDAmount_{t}\geq0\\
0, & TBDAmount_{t}<0
\end{cases}
\end{equation}
\begin{equation}
withdrawIND_{t}=\lnot\left(depositIND_{t}\right)
\end{equation}
$\lnot$ is the negation operators which changes 1 to 0 and 0 to 1
(also true to false and vice versa).
\item \label{enu:5-Calculate-the-new}Calculate the new dollar amounts that
each asset should hold (or the new capacity) based on the existing
investments and the new deployment amount. This is calculated as the
minimum, ideal and maximum amounts, $\left(minNewAmount_{it},idealNewAmount_{it},maxNewAmount_{it}\right)$,
that each asset $i$ can hold as follows,
\begin{align}
minNewAmount_{it} & =\min\left[\left(currentTotalAmount_{t}+TBDAmount_{t}\right)*minw_{it},\right.\\
 & \qquad\qquad\left.\left(currentTotalAmount_{t}+TBDAmount_{t}\right)*maxw_{it}\right]
\end{align}
\begin{align}
idealNewAmount_{it} & =\left(currentTotalAmount_{t}+TBDAmount_{t}\right)*idealw_{it}
\end{align}
\begin{align}
maxNewAmount_{it} & =\max\left[\left(currentTotalAmount_{t}+TBDAmount_{t}\right)*minw_{it},\right.\\
 & \qquad\qquad\left.\left(currentTotalAmount_{t}+TBDAmount_{t}\right)*maxw_{it}\right]
\end{align}
\item \label{enu:6-Across-each-asset,}Across each asset, $i$, calculate
the notional difference between the actual amount currently deployed
with the minimum and maximum new capacity, $minMaxCurrentDiff_{it}$.
This depends on whether we are deploying net new funds or withdrawing
funds.
\begin{align}
minMaxCurrentDiff_{it} & =\left(depositIND_{t}\right)*\left(maxNewAmount_{it}-currentAmount_{it}\right)\\
 & +\left(withdrawIND_{t}\right)*\left(minNewAmount_{it}-currentAmount_{it}\right)
\end{align}
\item \label{enu:7-Calculate-the-rebalance}Calculate the rebalance delta,
$rebalanceDelta_{it}$, on each asset, $i$, which indicates that
we might need to buy some assets even if there is a net withdrawal
and vice versa. This is to align the portfolio to stay within the
weight range due to the market movements. We also need the total rebalance
delta, $rebalanceDeltaTotal_{t}$, across the entire set of assets.
\begin{align}
rebalanceDelta_{it} & =\left(depositIND_{t}\right)*\min\left[\left(maxNewAmount_{it}-currentAmount_{it}\right),0\right]\\
 & +\left(withdrawIND_{t}\right)*\max\left[\left(minNewAmount_{it}-currentAmount_{it}\right),0\right]
\end{align}
\begin{equation}
rebalanceDeltaTotal_{t}=\sum_{i=1}^{k_{t}}\left(rebalanceDelta_{it}\right)
\end{equation}
\item \label{enu:8-Calculate,-,-on}Calculate, $rebalanceMinSizeDelta_{it}$,
on each asset, $i$, which indicates whether this asset rebalance
delta is less than the minimum order size, in which case this amount
will not get executed. We also need the total minimum size rebalance
delta, $rebalanceMinSizeDeltaTotal_{t}$, across the entire set of
assets.
\begin{align}
rebalanceMinSizeDelta_{it}=\begin{cases}
0, & \left|rebalanceDelta_{it}\right|\geq mins_{it}\\
rebalanceDelta_{it}, & \left|rebalanceDelta_{it}\right|<mins_{it}
\end{cases}
\end{align}
\begin{equation}
rebalanceMinSizeDeltaTotal_{t}=\sum_{i=1}^{k_{t}}\left(rebalanceMinSizeDelta_{it}\right)
\end{equation}
\item \label{enu:9-Across-each-asset,}Across each asset, $i$, set the
buy indicator, $buyIND_{it}=1$, which indicates if we need to buy
further shares on that asset. Otherwise, it will be set to zero and
indicates a sell order on that asset.
\begin{equation}
buyIND_{it}=\begin{cases}
1, & minMaxCurrentDiff_{it}\geq0\\
0, & minMaxCurrentDiff_{it}<0
\end{cases}
\end{equation}
\item \label{enu:10-Calculate-the-total}Calculate the total number of buy
and sell orders, $\left(totalBuyOrders_{t},totalSellOrders_{t}\right)$,
\begin{equation}
totalBuyOrders_{t}=\sum_{i=i}^{k_{t}}\left(buyIND_{it}\right)
\end{equation}
\begin{equation}
totalSellOrders_{t}=k_{t}-\sum_{i=i}^{k_{t}}\left(buyIND_{it}\right)
\end{equation}
\item \label{enu:11-Rank-the-capacity}Rank the capacity on each asset $i$,
such that the largest sell orders show up in the ranking first to
the smallest sell order. Then the largest buy order will show up until
the smallest buy order. This could simply be done by sorting the sell
orders from the largest absolute value (or minimum since they are
negative) to the smallest. Then sorting the buy orders from largest
to smallest and adding to the rank of each buy order the total number
of sell orders. This is also done by the following formula. 

$maxToMinRank_{it}$, $minToMaxRank_{it}$ and $assetCapRank_{it}$
indicate the rank when counting from the largest to the smallest (largest
buy order first since it is positive), smallest to the largest (largest
sell order first since it is negative) and the rank with sell orders
ranked before buy orders respectively. All three rankings give a unique
rank if more than two assets have the same sell or buy order size
by using $COUNT\left(\cdots\right)$, which counts the number of times
a number shows up in the entire asset capacity list.
\begin{align}
maxToMinRank_{it}= & RANK\left(minMaxCurrentDiff_{it},\text{HIGH-TO-LOW}\right)\\
+ & \quad COUNT\left(minMaxCurrentDiff_{it}\right)-1
\end{align}
\begin{align}
minToMaxRank_{it} & =RANK\left(minMaxCurrentDiff_{it},\text{LOW-TO-HIGH}\right)\\
 & +\quad COUNT\left(minMaxCurrentDiff_{it}\right)-1
\end{align}
\begin{align}
assetCapRank_{it} & =minToMaxRank_{it}\\
 & +buyIND_{it}*\left(maxToMinRank_{it}-minToMaxRank_{it}+totalSellOrders_{t}\right)
\end{align}

\item \label{enu:12-Then-calculate-the}Then calculate the capacity to fill,
$assetCapToFill_{it}$, on each asset $i$ based on the difference
between the current amount on each asset and maximum capacity, while
taking into account the total new amount to be deployed or withdrawn.
$assetRawCapFilled_{it}$, indicates how much capacity has been filled
without counting the capacity on asset $i$ and those ranked higher
than it by using the ranking, $assetCapRank_{it}$.
\begin{align}
assetRawCapFilled_{it} & =\sum_{i=1}^{\left(assetCapRank_{it}-1\right)}\left(minMaxCurrentDiff_{it}\right)
\end{align}
\begin{align}
assetCapToFill_{it} & =\left(depositIND_{t}\right)*\min\left[minMaxCurrentDiff_{it}\right.\\
 & \qquad\qquad\qquad,\left(TBDAmount_{t}+rebalanceMinSizeDeltaTotal_{t}\right.\\
 & \qquad\qquad\qquad-\min\left[\left(TBDAmount_{t}+rebalanceMinSizeDeltaTotal_{t}\right)\right.\\
 & \qquad\qquad\qquad,\left.\left.\left.assetRawCapFilled_{it}\right]\right)\right]\\
 & +\left(withdrawIND_{t}\right)*\max\left[minMaxCurrentDiff_{it}\right.\\
 & \qquad\qquad\qquad,\left(TBDAmount_{t}-rebalanceDeltaTotal_{t}\right.\\
 & \qquad\qquad\qquad-\max\left[\left(TBDAmount_{t}-rebalanceDeltaTotal_{t}\right)\right.\\
 & \qquad\qquad\qquad,\left.\left.\left.assetRawCapFilled_{it}\right]\right)\right]
\end{align}
Alternately, we can get $assetCapToFill_{it}$, on each asset $i$,
using multiple steps and with a few intermediate asset specific calculations
as shown below,
\begin{align}
assetBoundCapToFill_{it} & =\left(depositIND_{t}\right)*\min\left[\right.\\
 & \qquad\qquad\qquad\left(TBDAmount_{t}+rebalanceMinSizeDeltaTotal_{t}\right)\\
 & \qquad\qquad\qquad\qquad\qquad\left.,assetRawCapFilled_{it}\right]\\
 & +\left(withdrawIND_{t}\right)*\max\left[\left(TBDAmount_{t}-rebalanceDeltaTotal_{t}\right)\right.\\
 & \qquad\qquad\qquad\qquad\qquad\left.,assetRawCapFilled_{it}\right]
\end{align}
\begin{align}
assetCapInclusive_{it} & =\left(depositIND_{t}\right)*\min\left[\right.\\
 & \qquad\qquad\qquad\left(TBDAmount_{t}+rebalanceMinSizeDeltaTotal_{t}\right)\\
 & \qquad\qquad\qquad\left.,\left(minMaxCurrentDiff_{it}+assetBoundCapToFill_{it}\right)\right]\\
 & +\left(withdrawIND_{t}\right)*\max\left[\left(TBDAmount_{t}-rebalanceDeltaTotal_{t}\right)\right.\\
 & \qquad\qquad\qquad\left.,\left(minMaxCurrentDiff_{it}+assetBoundCapToFill_{it}\right)\right]
\end{align}
\begin{align}
assetCapToFill_{it} & =\left(depositIND_{t}\right)*\min\left[minMaxCurrentDiff_{it}\right.\\
 & \qquad\qquad\qquad\left.,\left(assetCapInclusive_{it}-assetBoundCapToFill_{it}\right)\right]\\
 & +\left(withdrawIND_{t}\right)*\max\left[minMaxCurrentDiff_{it}\right.\\
 & \qquad\qquad\qquad\left.,\left(assetCapInclusive_{it}-assetBoundCapToFill_{it}\right)\right]
\end{align}
\item \label{enu:13-Calculate-the-minimum}Calculate the minimum number
of orders, $minOrders_{it}$, on each asset $i$ and additional orders, 

$additionalOrders_{it}$, on each asset $i$ to get the total number
of orders, $totalOrders_{it}$, as shown below,
\begin{equation}
minOrders_{it}=\begin{cases}
1, & \left|assetCapToFill_{it}\right|\geq mins_{it}\\
0, & \left|assetCapToFill_{it}\right|<mins_{it}
\end{cases}
\end{equation}
\begin{equation}
additionalOrders_{it}=\left\lfloor \frac{\left|assetCapToFill_{it}\right|}{maxs_{it}}\right\rfloor 
\end{equation}
\begin{equation}
totalOrders_{it}=minOrders_{it}+additionalOrders_{it}
\end{equation}

\item \label{enu:14-Calculate-the-order}Calculate the order size, $orderSize_{it}$,
on each asset $i$, 
\begin{equation}
orderSize_{it}=\left(minOrders_{it}\right)*\left(\frac{assetCapToFill_{it}}{\left\lfloor \frac{\left|assetCapToFill_{it}\right|}{maxs_{it}}\right\rfloor +1}\right)
\end{equation}
\end{enumerate}
We have thus arrived at the number of orders on each asset and the
size of each order. Figures (\ref{fig:Rebalancing-Illustration:-Asset-Weights};
\ref{fig:Rebalancing:-Min-Max-Weights-and-Notional-Amounts}; \ref{fig:Rebalancing-Illustration:-Simple-Order-Schedule};
\ref{fig:Rebalancing-Illustration:-Cascading-Capacity-Fill}; \ref{fig:Rebalancing-Illustration:-Cascading-Order-Schedule})
show the order schedule based on the cascading rebalancing and also
several intermediate variables being calculated as required by Algorithm
(\ref{alg:Rebalancing-Algorithm}). The illustrations also provide
a comparison of cascading rebalancing versus a simple rebalancing
method - similar to what is practiced currently. 

As noticed, sell orders are always sent first and then the buy orders.
This is because once the sell orders are completed, the amount from
these sell orders are used to buy other assets. It is recommended
that we send the buy orders once we have got confirmation that the
sell orders have completed successfully. The amount that we receive
from the sell orders might be smaller and hence the amount to buy
needs to adjusted accordingly. Some sell orders might fail and these
errors should be handled and the buy amounts should be adjusted accordingly.
This is accomplished by changing, $rebalanceMinSizeDeltaTotal_{t}$,
by the difference between the value of sell orders when placed and
the value actually received. Then, we need to recalculate, $assetCapToFill_{it}$,
on each asset $i$.

It is also recommended that there be a delay of a few seconds between
orders. Lastly, something we need to consider for later, is to change
the size of the orders marginally, so that the orders do not look
similar and we cannot be front run or other possibilities from other
traders.

\section{\label{sec:Rebalancing-Flow-Chart}Rebalancing Flow Flow Chart}

\begin{figure}[H]
\includegraphics[width=18cm,height=10cm]{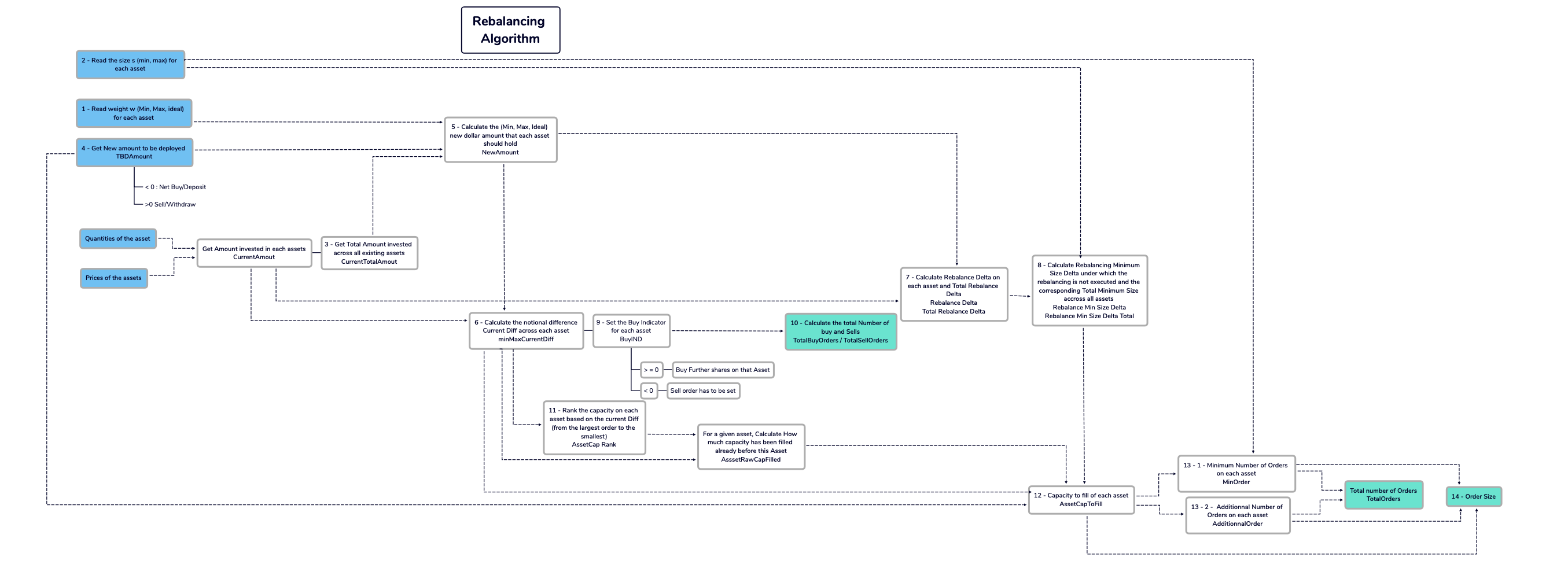}

\caption{Rebalancing Flow Flow Chart: Cascading Waterfall Round Robin Algorithm
\label{fig:Rebalancing-Flow-Flow}}
\end{figure}

The flow chart in Figure (\ref{fig:Rebalancing-Flow-Flow}) corresponds
to all the steps mentioned in Algorithm (\ref{alg:Rebalancing-Algorithm})
in Section (\ref{sec:The-Cascading-Waterfall-Rebalancing}).

\section{\label{sec:Gas-Fees-versus-Slippage}Alien vs. Predator, aka, Gas-Fees
versus Slippage: Coming Soon To Every Blockchain Network Near You}
\begin{itemize}
\item In this section we describe in more details the intuitive explanations
- regarding the minimum and maximum order size that depend on gas
fees and slippage - provided in Section (\ref{subsec:Blockchain-Based-Rebalancing})
to supplement the main algorithm described in Section (\ref{sec:The-Cascading-Waterfall-Rebalancing}).
\end{itemize}

\subsection{\label{subsec:Sequel-to-David}Prequel: David versus Goliath (You
against The Markets)}

The recent blockbuster book, David and Goliath: Underdogs, Misfits,
and the Art of Battling Giants (Gladwell 2013), talks about the advantages
of disadvantages, which in the legendary battle refers to (among other
things) the nimbleness that David possesses due to his smaller size
and lack of armor, that comes in handy while defeating the massive
and seemingly unbeatable Goliath. Despite the inspiring tone of the
story the efforts of the most valiant financial market participant
can seem puny and turn out to be inadequate, as it gets undone when
dealing with the gargantuan and mysterious temperament of uncertainty
in the markets. 

Another main feature of the David versus Goliath story is the tool
(sling: End-note \ref{enu:A-sling-is}) that David uses to defeat
Goliath. In this section, we hope to provide tools for market participants
to contend with the Goliath-like uncertainty in financial markets.
A trader’s conundrum is whether (and how much) to trade during a given
interval or wait for the next interval when the price momentum is
more favorable to his direction of trading. But given the nature of
uncertainty in the social sciences, any weapon might prove to be insufficient
compared to the sling that delivered the fatal blow to Goliath, until
perhaps, one can discern the ability to read the minds of all the
market participants. That being said, the techniques in this section
will go a long way towards helping participants and making their life
easier when confronting the markets. In addition, the mechanisms we
provide can be useful for combating uncertainty and aiding better
decision making in many areas of the social sciences.

Note that, Slippage is also known as Market Impact. The problem of
minimizing slippage by finding an optimal trade size is extensively
studied in traditional finance. Numerous models have been developed
that provide insights to market participants and deliver a well curated
trade execution schedule (Bertsimas \& Lo 1998; Almgren \& Chriss
2001; Karastilo 2020).

\subsection{\label{subsec:Trading-Block-Size}Checking Block-Busters and Calculating
Trading Block-Sizes}

The size of all the trades done on a blockchain network have to be
optimized so that gas fees and slippage are minimized. If we try to
decrease slippage, we will have to do more trades and hence incur
higher gas fees. Likewise, if we try to reduce gas fees, by trading
bigger executions, we will take on more market impact or slippage.

The trading block size calculations will have to be implemented in
a separate technological component. In the initial phases - for simplicity
- we can have these block sizes computed in a spreadsheet - or other
data store whenever necessary - and we can use it in the algorithm
in Section (\ref{sec:The-Cascading-Waterfall-Rebalancing}). Once
the rebalancing portion and weight calculations are performing satisfactorily
the focus can shift on rigorous implementation of the trading block
size calculations using live data. Optimization of the trading block
sizes are less important for now than the weight calculations and
rebalancing. A separate calculation procedure will calculate and output
the minimum and maximum size of trades across each asset into a data
store from which the rebalancing engine will access these values.
This size is related to the average daily volume across each asset,
average network gas fees and a few other factors. This mechanism which
is ideally implemented as an independent portion will need historical
market data and generate the block size list. $\left(mins_{it},maxs_{it}\right)$
are the minimum and maximum trading block sizes recommend for asset
$i$ at time $t$. In the initial iterations of the platform, the
following size calculations will suffice. These will be enhanced later
depending on various considerations.
\begin{itemize}
\item Input: Historical Market Data and Asset List
\item Output: Minimum and Maximum Trading Block Size
\begin{equation}
mins_{it}=\max\left[\left(AVGGASFEES_{it}\right)*\left(MINSIZEMULTIPLIER_{it}\right),MINSIZEPARAM{}_{it}\right]
\end{equation}
Here, $AVGGASFEES_{it}$, $MINSIZEMULTIPLIER_{it}$ and $MINSIZEPARAM_{it}$
represent respectively, at time $t$, the average gas fees on the
network that asset $i$ is part of, a factor to set the order size
as a multiple of the average gas fees and a safety parameter to ensure
that order sizes are not too small. $\left(MINSIZEMULTIPLIER_{t}=1000\right)$
is a suggested value such that gas fees (implicit transaction costs)
will be around 0.1\% of the order size. $MINSIZEPARAM_{it}=25000$
in USD terms is a suggested value; but it will generally have to be
different across networks. Note that $AVGGASFEES_{it}$, $MINSIZEMULTIPLIER_{it}$
and $MINSIZEPARAM_{it}$ will be the same across many assets. Some
indicative gas fees are given in Figure (\ref{fig:Network-Comparison}).
\begin{equation}
maxs_{it}=\min\left(\left[\frac{AVGDAILYVOLUME_{it}}{MAXSIZEDIVISOR_{it}}\right],\left[\frac{LIQUIDITYPOOLDEPTH_{it}}{2*\left(MAXSIZEDIVISOR_{it}\right)}\right],MAXSIZEPARAM_{it}\right)
\end{equation}
Here, $AVGDAILYVOLUME_{it}$, $LIQUIDITYPOOLDEPTH_{it}$, $MAXSIZEDIVISOR_{it}$
and 

$MAXSIZEPARAM_{it}$ represent respectively, at time $t$, the average
daily volume of asset $i$ on the centralized exchange to which orders
for that asset will be sent, the total depth of the liquidity pool
from which the asset will be procured or liquidated, a factor to set
the order size as a fraction of the daily volume or the liquidity
pool depth and a safety parameter to ensure that order sizes are not
too large. $\left(MAXSIZEDIVISOR_{it}=1000\right)$ is a suggested
value such that the market impact (implicit transaction costs) will
be around 0.1\% of the transaction costs. Cleary these parameters
will be revisited and estimated based on several factors using actual
historical data mixed with human instincts. $MAXSIZEPARAM_{it}=200000$
in USD terms is a suggested value; but it will generally have to be
different across networks and this default should be such that the
maximum slippage is less than 1 percent. Note that $MAXSIZEDIVISOR_{it}$
and $MAXSIZEPARAM_{it}$ will be the same across many assets. Clearly,
we can set $MAXSIZEDIVISOR_{it}$ to be different for centralized
exchanges and decentralized exchange liquidity pools.
\begin{figure}[H]
\includegraphics[width=18cm]{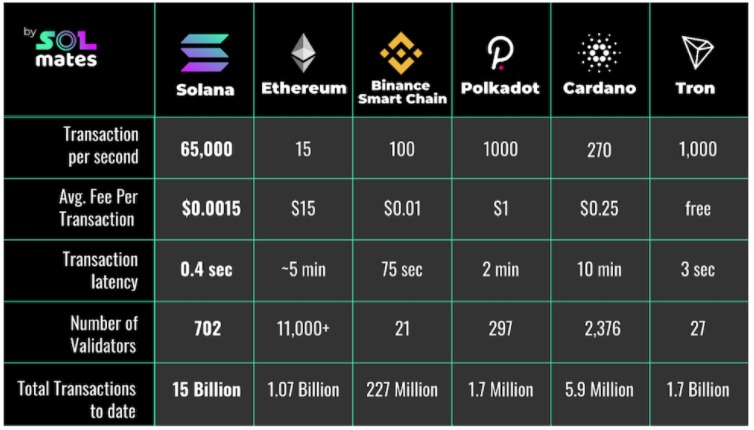}

Source: \href{https://chaindebrief.com/introduction-to-solana/}{Introduction To Solana,  Chaindebrief Article Link }\caption{\label{fig:Network-Comparison}Blockchain Network Comparison}
\end{figure}

\item Figure (\ref{fig:Rebalancing-Illustration:-Simple-Order-Schedule})
shows the minimum and maximum trade size being used - at the asset
level - as part of the rebalancing algorithm in Section (\ref{sec:The-Cascading-Waterfall-Rebalancing}).
\end{itemize}

\section{\label{sec:Weight-CE}Weight Calculation Engine: Obtaining Mathematical
Risk Parity}

Kashyap (2022) provides more detailed and intuitive explanations to
supplement the methodologies described here. The weight calculations
will have to be implemented in a separate technological component.
Initially - for simplicity - we can have these weights computed in
a spreadsheet - or other data store, on demand as necessary- and we
can use it in the algorithm in Section (\ref{sec:The-Cascading-Waterfall-Rebalancing}).

Within the first few weeks of operation, the goal will be to calculate
weights weekly or even several times during a week if possible. Once
the rebalancing portion is performing satisfactorily the focus can
shift to the weight calculation mechanism, which is extremely crucial,
and to have it connected to live data feeds. A separate calculation
routine will calculate and output the minimum, ideal and maximum weights
across each asset into a data store from which the rebalancing engine
will access these values. VVV (Velocity of Volatility and Variance)
weights will be the core of this engine and our reference weight.
This central portion will be mixed with other suitable adjustments
to get the overall weights on the assets. The rebalancing mechanism
which is ideally implemented as an independent portion will read these
weights from the data store and generate the order list. $\left(minw_{it},idealw_{it},maxw_{it}\right)$
are the minimum, ideal and maximum weights recommend for asset $i$
at time $t$.
\begin{itemize}
\item Input: Historical Market Data (Daily Open, Close, High, Low Prices)
and Asset List
\item Output: Minimum, Ideal and Maximum Weights
\end{itemize}
The following weight calculation methodologies will need to be implemented
initially. These can be modified to more sophisticated techniques
in subsequent iterations. We need the return, risk (volatility) across
each asset and the covariance between each asset pair to calculate
the weights for each asset, $i$, in the portfolio at time $t$. The
total number of assets in the portfolio is $k_{t}$ assets at time
$t$.

\subsection{\label{subsec:Return-and-Risk}Return and Risk (Volatility) Calculations}

The weight calculation depends on calculating the return of asset
prices on a daily basis (could be at other frequencies as well, but
we start with daily returns). The return will be the continuously
compounded measure. The volatility will be the standard deviation
of the continuously compounded returns over a historical time period,
generally around 90 days. The volatility will need to be calculated
on a rolling 90 days basis. 
\begin{enumerate}
\item \label{enu:Continously-Compounded-Return,}Continuously Compounded
Return, $R_{t}$, at any time $t$ is given by the logarithm of the
ratio of the Price at time $t$, $P_{t}$, and the Price at time $t-1$,
$P_{t-1}$, as shown below,
\begin{equation}
R_{t}=\ln\left(\frac{P_{t}}{P_{t-1}}\right)
\end{equation}
\item \label{enu:Volatility-at-time}Volatility at time $t$, $\sigma_{t}$
is calculated as below using the average return, $\bar{R}_{t}$, at
time $t$ and then calculating the standard deviation. It can be done
directly if suitable libraries are available. $\sigma_{t}^{2}$ is
the variance at time $t$. Here, $T=90$ unless otherwise stated.
\begin{equation}
\bar{R}_{t}=\left(\frac{1}{T}\right)\sum_{i=t}^{t-T}R_{i}\label{eq:Return-Estimate}
\end{equation}
\begin{equation}
\sigma_{t}^{2}=\left(\frac{1}{T-1}\right)\sum_{i=t}^{t-T}\left(R_{i}-\bar{R}_{t}\right)^{2}
\end{equation}
\begin{equation}
\sigma_{t}=\sqrt{\left(\frac{1}{T-1}\right)\sum_{i=t}^{t-T}\left(R_{i}-\bar{R}_{t}\right)^{2}}\label{eq:Volatility-Risk-Estimate}
\end{equation}
\item Calculate the volatility for the longest historical time period possible
for each asset. Possibly for the last 360 days or longer. Some new
assets might have a relatively shorter price time series, and it is
okay to calculate volatility for those assets for the number of days
for which prices are available.
\end{enumerate}

\subsection{\label{subsec:Asset-Weight-Calculation}Asset Weight Calculations}
\begin{enumerate}
\item \label{enu:Equal-Weighted-Scenario}Equal Weighted Scenario gives
the same weight, $wequal_{it}$, for each asset, $i$, in the portfolio
at time $t$. The total number of assets in the portfolio is $k_{t}$
assets at time $t$. This is a fairly robust strategy and performs
well under many scenarios and is highly tolerant to estimation errors.
\begin{equation}
wequal_{it}=\frac{1}{k_{t}}
\end{equation}
\item \label{enu:Simple-Variance-Weighted}Simple Variance Weighted Scenario
gives the weight, $wsimplevar_{it}$, for each asset, $i$, in the
portfolio at time $t$. The total number of assets in the portfolio
is $k_{t}$ assets at time $t$. This weighting scheme can be shown
to give the lowest variance of the portfolio under certain conditions.
\begin{equation}
wsimplevar_{it}=\left(\frac{\frac{1}{\sigma_{it}^{2}}}{\sum_{j=1}^{k_{t}}\frac{1}{\sigma_{jt}^{2}}}\right)
\end{equation}
Here, $\sigma_{it}^{2}$ is the variance of asset, $i$, at time $t$,
calculated using Steps (\ref{enu:Continously-Compounded-Return,};
\ref{enu:Volatility-at-time}) in Section (\ref{subsec:Return-and-Risk}).
\item \label{enu:Simple-Parity-Weighted}Simple Parity Weighted Scenario
gives the weight, $wsimpleparity_{it}$, for each asset, $i$, in
the portfolio at time $t$. The total number of assets in the portfolio
is $k_{t}$ assets at time $t$. This weighting scheme holds when
correlations tend to one, which is a reasonable assumption during
market crashes. When we set the weights to be inverse of the volatilities
divided by the sum of all the inverse volatilities, we get one solution
such that the risk contribution from each asset is equal. This solution
also satisfies the constraints that the sum of the weights is equal
to one and the weights are positive.
\begin{equation}
wsimpleparity_{it}=\left(\frac{\frac{1}{\sigma_{it}}}{\sum_{j=1}^{k_{t}}\frac{1}{\sigma_{jt}}}\right)
\end{equation}
Here, $\sigma_{it}$ is the volatility of asset, $i$, at time $t$,
calculated using Steps (\ref{enu:Continously-Compounded-Return,};
\ref{enu:Volatility-at-time}) in Section (\ref{subsec:Return-and-Risk}).
\item \label{enu:Risk-Parity-Weighted}Risk Parity Weighted Scenario gives
the weight, $wriskparity_{it}$, for each asset, $i$, in the portfolio
at time $t$. The total number of assets in the portfolio is $k$
assets at time $t$. We remove the time suffix for the number of assets
to lighten the notation. This weighting schemes ensures that the risk
contribution of each asset towards the overall portfolio risk is equal.
The covariance between the assets is also considered in this formulation.
Using matrix notation, we can write the variance of the portfolio
as,
\begin{equation}
\boldsymbol{\sigma_{p}\left(w\right)^{2}}=\boldsymbol{w'Xw}\label{eq:Matrix_Regression-No-Constraint}
\end{equation}
$\boldsymbol{\sigma_{p}^{2}}$ is the portfolio variance which is
expressed using the covariance matrix, $\boldsymbol{X}$ of the $k$
individual securities. $\boldsymbol{w}$ is the weight vector to be
estimated. One condition on the weights can be that that $\boldsymbol{w'1}=1$
Here, $\boldsymbol{1}$ is the column vector (or k{*}1 matrix) with
1. Simplifying this further, using the condition that the risk contribution
from each asset is equal, we arrive at the below formulation, which
can be solved using fixed point techniques,
\begin{equation}
\boldsymbol{w_{i}}=\frac{\boldsymbol{\sigma_{p}\left(w\right)^{2}}}{\boldsymbol{k\left(\sum w\right)_{i}}}
\end{equation}
This is equivalently solved as the below minimization problem,
\begin{equation}
\underset{\boldsymbol{w}}{\min}\sum_{i=1}^{k}\left(\boldsymbol{w_{i}}-\frac{\boldsymbol{\sigma_{p}\left(w\right)^{2}}}{\boldsymbol{k\left(\sum w\right)_{i}}}\right)
\end{equation}

A python library to solve the above problem, and to calculate the
risk Parity weights, is available at this location: (\href{https://github.com/dppalomar/riskParityPortfolio}{Risk Parity Portfolio,  Python Library}).
The corresponding R package is available at this location: (\href{https://cran.r-project.org/web/packages/riskParityPortfolio/vignettes/RiskParityPortfolio.html}{Risk Parity Portfolio,  R Library};
\href{https://cran.r-project.org/web/packages/riskParityPortfolio/riskParityPortfolio.pdf}{Risk Parity Portfolio,  R Reference PDF}).
\item \label{enu:Velocity-of-Volatility}Velocity of Volatility and Variance
or Volatility of Volatility and Variance (VVV) Scenario gives the
weight, $wvvv_{it}$, for each asset, $i$, in the portfolio at time
$t$. The total number of assets in the portfolio is $k_{t}$ assets
at time $t$. In this scenario, we adjust the volatilities higher
by the volatility of the volatilities of each asset. This weighting
scheme holds when correlations tend to one, which is a reasonable
assumption during market crashes. When we set the weights to be inverse
of the adjusted volatilities divided by the sum of all the inverse
adjusted volatilities, we get one solution such that the risk contribution
from each asset is equal. This solution also satisfies the constraints
that the sum of the weights is equal to one and the weights are positive.
We first use the formulae in Steps (\ref{enu:Continously-Compounded-Return,};
\ref{enu:Volatility-at-time}) in Section (\ref{subsec:Return-and-Risk}),
along with similar notation, to calculate the volatility of volatility
as follows,
\begin{equation}
V_{it}=\ln\left(\frac{\sigma_{it}}{\sigma_{i,t-1}}\right)
\end{equation}
\begin{equation}
\bar{V}_{it}=\left(\frac{1}{T}\right)\sum_{l=t}^{t-T}V_{il}\label{eq:Volatility-Change-Estimate}
\end{equation}
\begin{equation}
\sigma_{it}^{VVVFactor}=\sqrt{\left(\frac{1}{T-1}\right)\sum_{l=t}^{t-T}\left(V_{il}-\bar{V}_{it}\right)^{2}}\label{eq:Volatility-Volatility-Estimate}
\end{equation}
\begin{equation}
\sigma_{it}^{VVV}=\sigma_{it}+\left(\theta\right)\sigma_{it}^{VVVFactor}\label{eq:Volatility-Plus-VVVFactor}
\end{equation}
Here, $\theta$, is a parameter that has to be calibrated to give
the length of time a market downturn occurs. We set $\theta=1$ for
now and we can consider further estimates of $\theta$ in later iterations.
\begin{equation}
wvvv_{it}=\left(\frac{\frac{1}{\sigma_{it}^{VVV}}}{\sum_{j=1}^{k_{t}}\frac{1}{\sigma_{jt}^{VVV}}}\right)
\end{equation}
The ideal weight $idealw_{it}$ mentioned in Algorithm (\ref{alg:Rebalancing-Algorithm})
corresponds to the weights calculated using this procedure.
\item \label{enu:Covariance-Weighted-Scenario}Covariance Weighted Scenario
gives the weight, $wcovar_{it}$, for each asset, $i$, in the portfolio
at time $t$. The total number of assets in the portfolio is $k_{t}$
assets at time $t$. We remove the time suffix for the number of assets
to lighten the notation. These are the traditional Markowitz inspired
portfolio theory weights. The weight, $wcovar_{it}$ for each asset,
$i$ is given by using the corresponding row in the weight vector
derived below. Consider a portfolio variance optimization problem,
shown using matrix notation, in Equation (\ref{eq:Matrix_Regression}).
$\boldsymbol{\sigma_{p}^{2}}$ is the portfolio variance which is
expressed using the covariance matrix, $\boldsymbol{X}$ of the $k$
individual securities. $\boldsymbol{w}$ is the weight vector to be
estimated with the condition that $\boldsymbol{w'1}=1$. Here, $\boldsymbol{1}$
is the column vector (or k{*}1 matrix) with 1.
\begin{equation}
\boldsymbol{\sigma_{p}^{2}}=\boldsymbol{w'Xw}\qquad;\boldsymbol{w'1}=1\label{eq:Matrix_Regression}
\end{equation}
Minimizing variance using Lagrangian multiplier techniques gives the
following weights, 
\begin{equation}
\boldsymbol{w}=\frac{\boldsymbol{X^{-1}1}}{\boldsymbol{1'X^{-1}1}}
\end{equation}
Note that several alternate constrained optimizations are possible
and there is a detailed discussion of the issues that arise with constrained
optimizations in Kashyap (2022). The limitations of optimization methodologies
and the need for range based techniques, which introduce randomness
in the decision process are discussed in detail in the series, ``Fighting
Uncertainty with Uncertainty'': (End-note \ref{enu:The-range-based};
Kabeer 2016). 
\item To get a weight range for each asset, we need a minimum and maximum
weight. The minimum and maximum weights for each asset will be arrived
at using the below formulae. 
\begin{equation}
minweight_{it}=\max\left[\min\left(All\:Weights\:For\:Asset\:i\right),\min\left(wvvv_{it},MINASSETWEIGHT_{t}\right)\right]
\end{equation}
\begin{equation}
maxweight_{it}=\min\left[\max\left(All\:Weights\:For\:Asset\:i\right),\max\left(wvvv_{it},MAXASSETWEIGHT_{t}\right)\right]
\end{equation}
Here, $MINASSETWEIGHT_{t}$ and $MAXASSETWEIGHT_{t}$ are based on
risk management guidelines which will also be reviewed and revised
continuously. Suggested values to start with are $MINASSETWEIGHT_{t}=0.0$
and $MAXASSETWEIGHT_{t}=0.15$. We will consider several other weight
calculation techniques in later stages. Further fine tuning of the
range for each asset will be based on these additional weight calculations,
observed empirical performance and other additional considerations.
The minimum and maximum weights $\left(minw_{it},maxw_{it}\right)$
mentioned in Algorithm (\ref{alg:Rebalancing-Algorithm}) correspond
to the weights calculated using this procedure.
\item An initial simplification can be to calculate weights on each network
separately. This is related to limitations on bridge capacities and
related bottlenecks (Kashyap 2023). This means that we will have weights
optimized for each network. Hence, instead of relying on global portfolio
weights and arriving at network weights from the global portfolio
weights, we will calculate weights on each network by considering
it to be a global portfolio.
\item Figures (\ref{fig:Rebalancing-Illustration:-Asset-Weights}; \ref{fig:Rebalancing:-Min-Max-Weights-and-Notional-Amounts})
show weights being calculated using different techniques and how we
arrive at the minimum and maximum capacity - at the asset level -
as part of the rebalancing algorithm in Section (\ref{sec:The-Cascading-Waterfall-Rebalancing}).
Several other weighting mechanisms are possible and much further research
can be conducted in this domain - seeking to improve portfolio performance
- and can be implemented progressively at conducive times. These additional
weighting techniques can either replace or can be included as an extra
weight before choosing the minimum and maximum for each asset.
\end{enumerate}

\section{\label{sec:Numerical-Results}Numerical Results}

\begin{figure}[H]
\includegraphics[width=18cm]{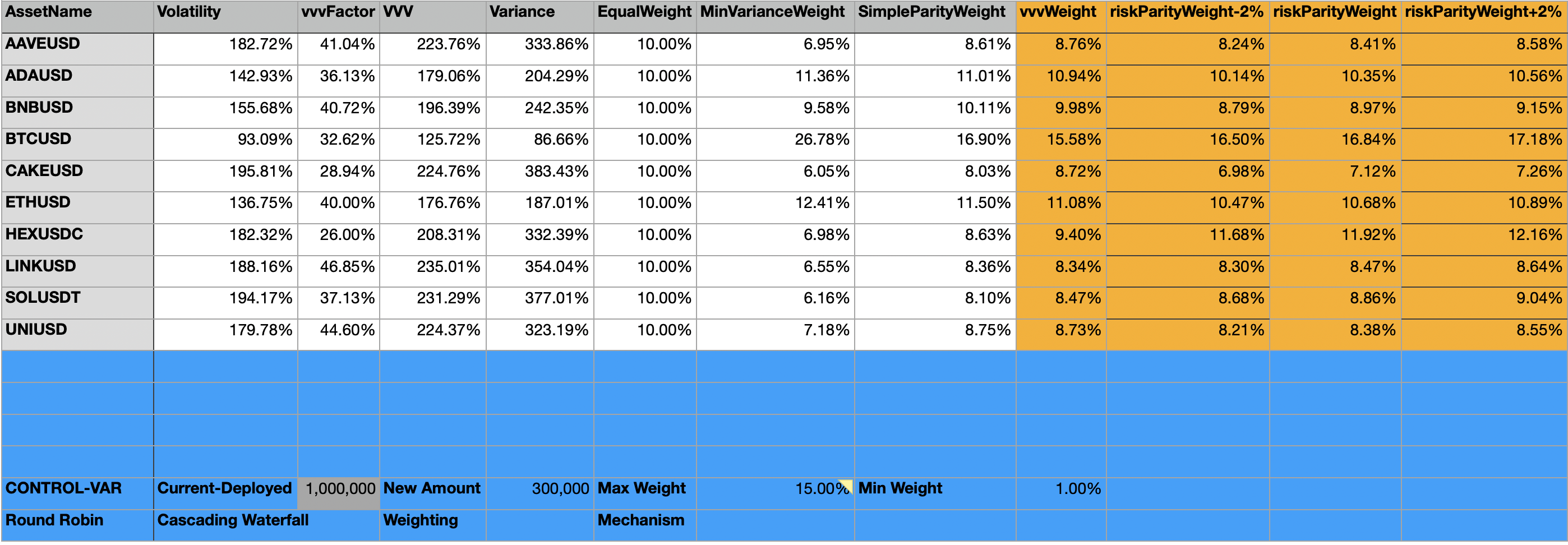}\caption{\label{fig:Rebalancing-Illustration:-Asset-Weights}Rebalancing Illustration:
Asset Weights}
\end{figure}

Each of the tables in this section are referenced in the main body
of the article. Below, we provide supplementary descriptions for each
table. 
\begin{itemize}
\item The Table in Figure (\ref{fig:Rebalancing-Illustration:-Asset-Weights})
shows numerical examples related to the weight calculations described
in Section (\ref{sec:Weight-CE}) and the material in Section (\ref{subsec:Asset-Weight-Calculation}).
The numbers in the figure also show the volatility and variance at
the asset level. Certain control parameters are also given such as
the net inflow or outflow into the fund. We have a spreadsheet, which
can be shared upon request, wherein the control variables can be changed
to see how the cascading algorithm performs - by observing several
intermediate variables that are calculated - and compare it to a simple
rebalancing rule. 
\item The asset weights shown in the illustrations are based on real historical
asset prices. The full data sample consists of daily observations
from October 31, 2020 to October 31, 2021. The asset names used are
based on naming conventions followed by most crypto data providers
(End-note \ref{enu:CoinMarketCap}; \ref{enu:Crypto-Ranking}). The
USD suffix in the asset name indicates that the prices were denominated
in US Dollars. The asset prices are expressed in stable coins such
as USDT or USDC which are pegged to the US dollar (Ante, Fiedler \&
Strehle 2021; End-note \ref{enu:Stablecoin}).
\item The simple rebalancing mechanism considers the difference between
the ideal notional that an asset should hold and the amount it currently
holds to arrive at the whether buy or sell orders need to be made.
In contrast, the cascading rebalancing mechanism considers the different
between the minimum notional that an asset should hold - or the maximum
notional amount an asset can hold depending on whether there is a
net deposit or withdrawal from the portfolio - and the amount it currently
holds to arrive at the whether buy or sell orders need to be made.
Numerous simulation based comparisons can be included either in appendix
of this paper or will be provided in a subsequent paper that will
have more analytical estimates and performance results.
\item The columns in Figure (\ref{fig:Rebalancing-Illustration:-Asset-Weights})
represent the following information respectively:
\begin{enumerate}
\item \textbf{AssetName} is the symbol or asset name that is part of the
portfolio being rebalanced. 
\item \textbf{Volatility} is the volatility of the corresponding asset. 
\item \textbf{vvvFactor} corresponds to the calculation of asset weights
based on the Velocity of Volatility and Variance technique described
in Section (\ref{subsec:Asset-Weight-Calculation}). 
\item \textbf{VVV} refers to the VVV adjusted volatility of the asset. 
\item \textbf{Variance} is the variance of the corresponding asset. 
\item \textbf{EqualWeight} is the weight allocated to the asset based on
the equal weighting technique. This is discussed in Point (\ref{enu:Equal-Weighted-Scenario})
in Section (\ref{subsec:Asset-Weight-Calculation}).
\item \textbf{MinVarianceWeight} is the weight allocated to the asset such
that the variance of the portfolio is minimized. This is discussed
in Point (\ref{enu:Covariance-Weighted-Scenario}) in Section (\ref{subsec:Asset-Weight-Calculation}).
\item \textbf{SimpleParityWeight} is the weight allocated to the asset based
on the simple parity scheme. This is discussed in Point (\ref{enu:Simple-Parity-Weighted})
in Section (\ref{subsec:Asset-Weight-Calculation}).
\item \textbf{vvvWeight} is the weight allocated to the asset based on the
VVV weighting technique. This is discussed in Point (\ref{enu:Velocity-of-Volatility})
in Section (\ref{subsec:Asset-Weight-Calculation}).
\item \textbf{riskParityWeight} is the weight allocated to the asset based
on the risk parity weighting technique. This is discussed in Point
(\ref{enu:Risk-Parity-Weighted}) in Section (\ref{subsec:Asset-Weight-Calculation}).
\item \textbf{riskParityWeight-2\%} is 2\% less than the weight in the risk
parity weighting technique.
\item \textbf{riskParityWeight+2\%} is 2\% more than the weight in the risk
parity weighting technique.
\end{enumerate}
\end{itemize}

\begin{figure}[H]
\includegraphics[width=18cm]{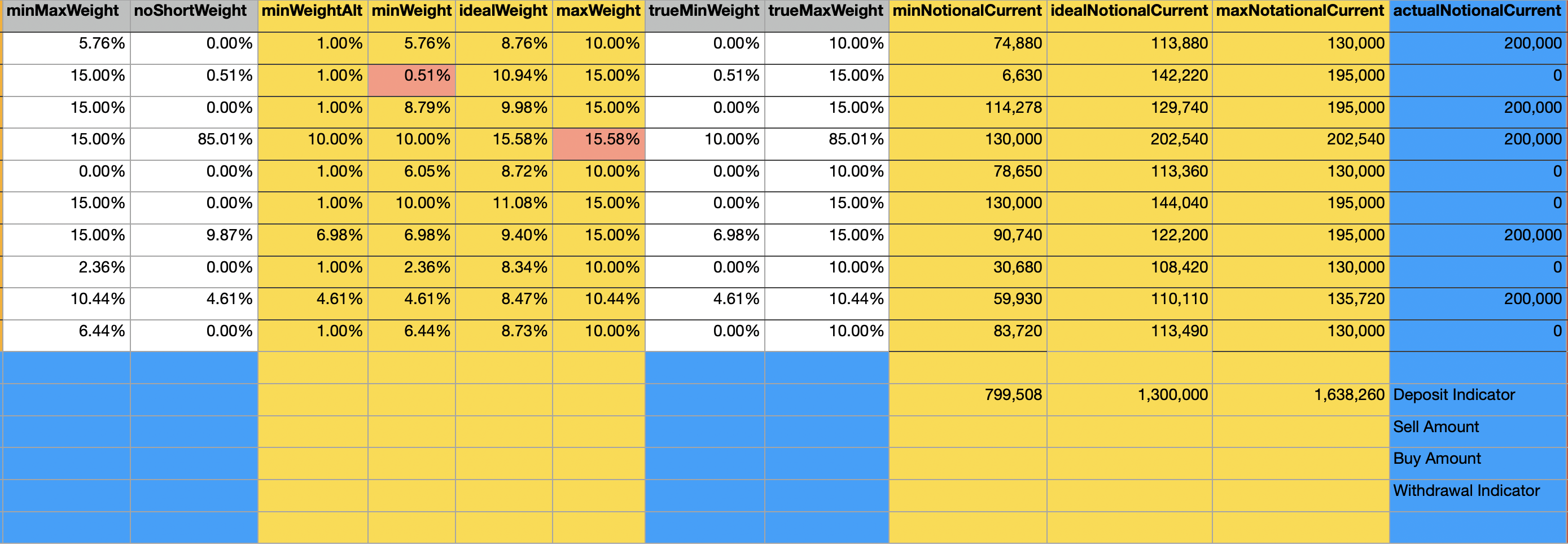}

\caption{\label{fig:Rebalancing:-Min-Max-Weights-and-Notional-Amounts}Rebalancing
Illustration: Current Notional Amounts}
\end{figure}

\begin{itemize}
\item The Table in Figure (\ref{fig:Rebalancing:-Min-Max-Weights-and-Notional-Amounts})
shows numerical examples related to how we arrive at the minimum and
maximum capacity for each asset, which is expressed in terms of the
notional amounts each asset can hold. The difference between asset
capacities and the current notional amounts deployed across each asset
- also given in the illustration - determine whether an asset will
receive new funds or funds will be withdrawn when there are net inflows
and outflows into the fund.
\item The columns in Figure (\ref{fig:Rebalancing:-Min-Max-Weights-and-Notional-Amounts})
represent the following information respectively: 
\begin{enumerate}
\item \textbf{minMaxWeight} is the weight allocated to the asset such that
the variance of the portfolio is minimized but where there are minimum
and maximum weight constraints set for each asset. The variance minimization
of the portfolio and the corresponding method to derive asset weights
is discussed in Point (\ref{enu:Covariance-Weighted-Scenario}) in
Section (\ref{subsec:Asset-Weight-Calculation}). The minimum and
maximum constraint across each asset is set at 0\% and 15\% respectively.
\item \textbf{noShortWeight} is the weight allocated to the asset such that
the variance of the portfolio is minimized but when no shorts are
allowed across any asset. This corresponds to not having a maximum
weight constraint on any asset but having a zero percent minimum level.
The variance minimization of the portfolio and the corresponding method
to derive asset weights is discussed in Point (\ref{enu:Covariance-Weighted-Scenario})
in Section (\ref{subsec:Asset-Weight-Calculation}). 
\item \textbf{minWeightAlt} is the minimum weight based on the minimum and
maximum weight bounds given in Figure (\ref{fig:Rebalancing-Illustration:-Asset-Weights}).
This weight is obtained after selecting the minimum and maximum based
on the weighting techniques discussed in Section (\ref{subsec:Asset-Weight-Calculation})
and Figure (\ref{fig:Rebalancing-Illustration:-Asset-Weights}). 
\item \textbf{minWeight} is the minimum weight based on selecting the minimum
across the various weighting techniques discussed in Section (\ref{subsec:Asset-Weight-Calculation})
and Figure (\ref{fig:Rebalancing-Illustration:-Asset-Weights}). 
\item \textbf{idealWeight} is obtained using the VVV weighting technique.
This is discussed in Point (\ref{enu:Velocity-of-Volatility}) in
Section (\ref{subsec:Asset-Weight-Calculation}).
\item \textbf{maxWeight} is the maximum weight based on selecting the maximum
across the various weighting techniques discussed in Section (\ref{subsec:Asset-Weight-Calculation})
and Figure (\ref{fig:Rebalancing-Illustration:-Asset-Weights}). 
\item \textbf{trueMinWeight} is the minimum weight without using the minimum
weight bound given in Figure (\ref{fig:Rebalancing-Illustration:-Asset-Weights}). 
\item \textbf{minNotionalCurrent} is the minimum notional capacity of the
asset based on the minimum weight of the corresponding asset and the
total current amount invested in the portfolio plus net deposits or
withdrawals to be made across all investors.
\item \textbf{idealNotionalCurrent} is the ideal notional capacity of the
asset based on the ideal weight of the corresponding asset and the
total current amount invested in the portfolio plus net deposits or
withdrawals to be made across all investors.
\item \textbf{maxNotionalCurrent} is the maximum notional capacity of the
asset based on the maximum weight of the corresponding asset and the
total current amount invested in the portfolio plus net deposits or
withdrawals to be made across all investors.
\item \textbf{actualNotionalCurrent} is the actual amount invested into
the corresponding asset.
\end{enumerate}
\end{itemize}

\begin{figure}[H]
\includegraphics[width=18cm]{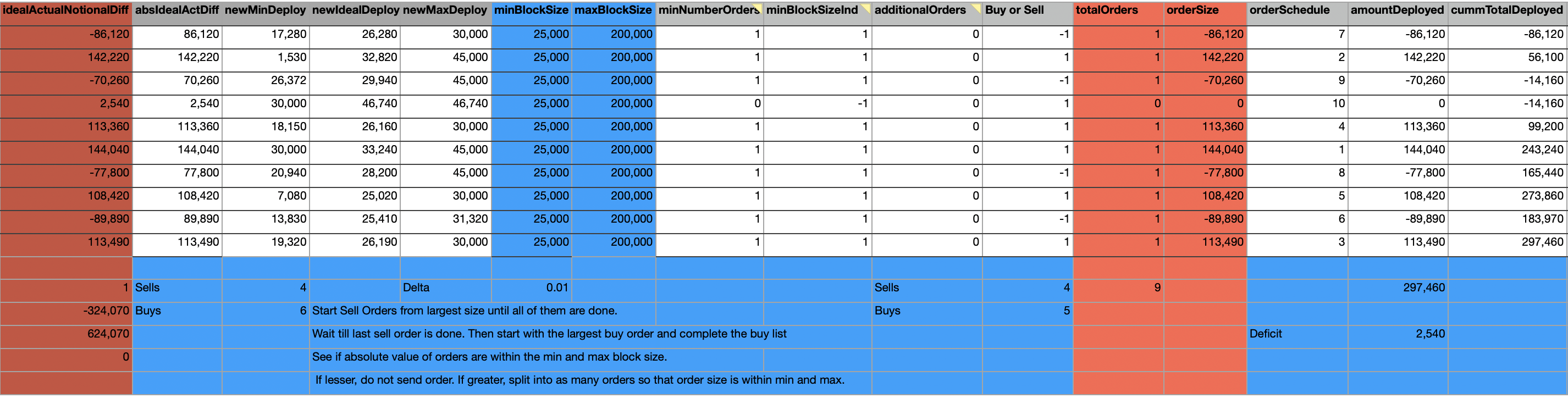}

\caption{\label{fig:Rebalancing-Illustration:-Simple-Order-Schedule}Rebalancing
Illustration: Simple Mechanism Order Schedule}
\end{figure}

\begin{itemize}
\item The Table in Figure (\ref{fig:Rebalancing-Illustration:-Simple-Order-Schedule})
shows numerical examples related to the minimum and maximum trade
size calculations described in Section (\ref{subsec:Trading-Block-Size})
based on the material in Section (\ref{sec:Gas-Fees-versus-Slippage})
regarding the trade-off between gas fees and trading slippage. This
figure also illustrates the order schedule based on a simple rebalancing
rule when trades are generated as assets move away from their intended
weights.
\item The columns in Figure (\ref{fig:Rebalancing-Illustration:-Simple-Order-Schedule})
represent the following information respectively: 
\begin{enumerate}
\item \textbf{idealActualNotionalDiff} is the difference between the ideal
notional capacity of the asset and the actual amount invested in the
asset.
\item \textbf{absIdealActDiff} is the difference between the ideal notional
capacity of the asset and the actual amount invested in the asset.
\item \textbf{newMinDeploy} is the minimum notional amount that should be
deployed - or invested - into this asset based on the minimum weight
of the corresponding asset and the net deposits or withdrawals to
be made across all investors.
\item \textbf{newIdealDeploy} is the ideal notional amount that should be
deployed - or invested - into this asset based on the ideal weight
of the corresponding asset and the net deposits or withdrawals to
be made across all investors.
\item \textbf{newMaxDeploy} is the maximum notional amount that should be
deployed - or invested - into this asset based on the maximum weight
of the corresponding asset and the net deposits or withdrawals to
be made across all investors.
\item \textbf{minBlockSize} is the minimum trade size for this asset based
on the calculations described in Section (\ref{subsec:Trading-Block-Size}).
\item \textbf{maxBlockSize} is the maximum trade size for this asset based
on the calculations described in Section (\ref{subsec:Trading-Block-Size}).
\item \textbf{minNumberOrders} is the minimum number of orders that need
to be made on this asset based on the simple rebalancing mechanism
described above Figure (\ref{fig:Rebalancing-Illustration:-Asset-Weights}).
This column shows zero - 0- if the absolute Value of difference between
ideal and actual notional is less than the minimum block size; and
1 otherwise. 
\item \textbf{minBlockSizeInd} indicates that the trade size on this asset
is below the minimum trade size for that asset. This column shows
minus one, -1, if the absolute Value of the difference between ideal
and actual notional is less than minimum block size; 1 otherwise. 
\item \textbf{additionalOrders} is the number of additional orders required
to fill the total deficit in this asset. This is the number of orders
in addition to the minimum number of orders indicated by \textbf{minNumberOrders.}
\item \textbf{Buy or Sell} indicates if we are making buy or sell trades
on this asset. 
\item \textbf{totalOrders} indicates the total number of orders to be made
on this asset. 
\item \textbf{orderSize} is the size of each trade to be made on this asset. 
\item \textbf{orderSchedule} indicates the sequence in which trades for
this asset will have to be executed in comparison to the other assets
in the portfolio.
\item \textbf{amountDeployed} indicates the total amount deployed on this
asset in this rebalancing event. This is the product of \textbf{totalOrders
}and\textbf{ orderSize.}
\item \textbf{cummTotalDeployed} indicates the cumulative total deployed
- summing up the amounts invested or withdrawn, bought or sold - across
all the assets in the portfolio starting from the asset in the first
row.
\end{enumerate}
\end{itemize}

\begin{figure}[H]
\includegraphics[width=18cm]{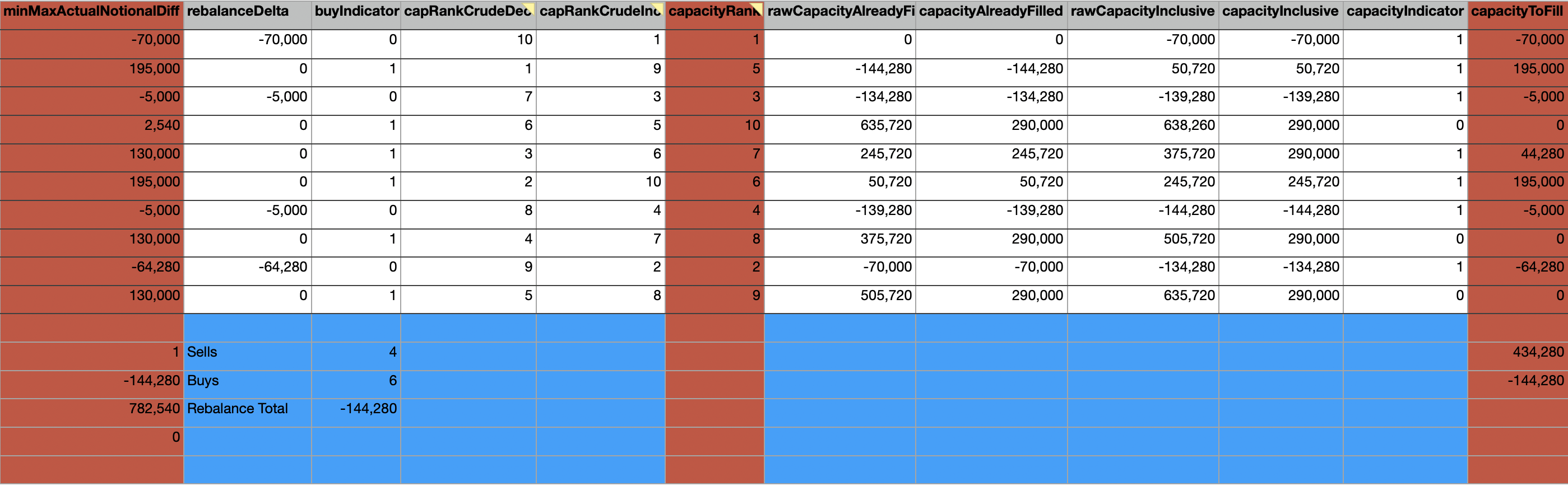}

\caption{\label{fig:Rebalancing-Illustration:-Cascading-Capacity-Fill}Rebalancing
Illustration: Cascading Mechanism Capacity To Fill}
\end{figure}

\begin{itemize}
\item The Table in Figure (\ref{fig:Rebalancing-Illustration:-Cascading-Capacity-Fill})
shows intermediate variables related to the calculations based on
the material described in Section (\ref{sec:The-Cascading-Waterfall-Rebalancing})
corresponding to the steps in Algorithm (\ref{alg:Rebalancing-Algorithm}).
This figure also gives a capacity ranking of the assets depending
on how much funds they can hold. Assets that need to reduce their
holdings - or those that generate sell orders - are higher up in the
capacity list and then assets which require inflows show up next -
both in decreasing absolute value of dollar amounts.
\item The columns in Figure (\ref{fig:Rebalancing-Illustration:-Cascading-Capacity-Fill})
represent the following information respectively: 
\begin{enumerate}
\item \textbf{minMaxActualNotionalDiff} is the difference between either
the minimum or maximum notional capacity of the asset and the actual
amount invested in the asset. This is explained in Step (\ref{enu:6-Across-each-asset,})
of Algorithm (\ref{alg:Rebalancing-Algorithm}) in Section (\ref{sec:The-Cascading-Waterfall-Rebalancing}). 
\item \textbf{rebalanceDelta} indicates that for the corresponding asset
there is a need add to the existing position even if there is a net
withdrawal and vice versa. This is explained in Steps (\ref{enu:7-Calculate-the-rebalance};
\ref{enu:8-Calculate,-,-on}) of Algorithm (\ref{alg:Rebalancing-Algorithm})
in Section (\ref{sec:The-Cascading-Waterfall-Rebalancing}). 
\item \textbf{buyIndicator} indicates that we need to buy additional positions
for this asset. This is explained in Step (\ref{enu:9-Across-each-asset,})
of Algorithm (\ref{alg:Rebalancing-Algorithm}) in Section (\ref{sec:The-Cascading-Waterfall-Rebalancing}). 
\item \textbf{capRankCrudeDec} indicates the ranking for this asset based
on capacity while comparing with all the other assets in the portfolio
in decreasing order. This is explained in Step (\ref{enu:11-Rank-the-capacity})
of Algorithm (\ref{alg:Rebalancing-Algorithm}) in Section (\ref{sec:The-Cascading-Waterfall-Rebalancing}). 
\item \textbf{capRankCrudeDec} indicates the ranking for this asset based
on capacity while comparing with all the other assets in the portfolio
in increasing order. This is explained in Step (\ref{enu:11-Rank-the-capacity})
of Algorithm (\ref{alg:Rebalancing-Algorithm}) in Section (\ref{sec:The-Cascading-Waterfall-Rebalancing}). 
\item \textbf{capacityRank} indicates the ranking for this asset based on
capacity while comparing with all the other assets in the portfolio
with the largest sells in decreasing order first and then largest
buys in decreasing order next. This is explained in Step (\ref{enu:11-Rank-the-capacity})
of Algorithm (\ref{alg:Rebalancing-Algorithm}) in Section (\ref{sec:The-Cascading-Waterfall-Rebalancing}). 
\item \textbf{rawCapacityAlreadyFilled} is an intermediate variable used
to arrive at the capacity to fill - \textbf{capacityToFill} - on that
asset. This is explained in Step (\ref{enu:12-Then-calculate-the})
of Algorithm (\ref{alg:Rebalancing-Algorithm}) in Section (\ref{sec:The-Cascading-Waterfall-Rebalancing}). 
\item \textbf{capacityAlreadyFilled} is an intermediate variable used to
arrive at the capacity to fill - \textbf{capacityToFill} - on that
asset. This is explained in Step (\ref{enu:12-Then-calculate-the})
of Algorithm (\ref{alg:Rebalancing-Algorithm}) in Section (\ref{sec:The-Cascading-Waterfall-Rebalancing}). 
\item \textbf{rawCapacityInclusive} is an intermediate variable used to
arrive at the capacity to fill - \textbf{capacityToFill} - on that
asset. This is explained in Step (\ref{enu:12-Then-calculate-the})
of Algorithm (\ref{alg:Rebalancing-Algorithm}) in Section (\ref{sec:The-Cascading-Waterfall-Rebalancing}). 
\item \textbf{capacityInclusive} is an intermediate variable used to arrive
at the capacity to fill - \textbf{capacityToFill} - on that asset.
This is explained in Step (\ref{enu:12-Then-calculate-the}) of Algorithm
(\ref{alg:Rebalancing-Algorithm}) in Section (\ref{sec:The-Cascading-Waterfall-Rebalancing}). 
\item \textbf{capacityIndicator} is an intermediate variable used to arrive
at the capacity to fill - \textbf{capacityToFill} - on that asset.
This is explained in Step (\ref{enu:12-Then-calculate-the}) of Algorithm
(\ref{alg:Rebalancing-Algorithm}) in Section (\ref{sec:The-Cascading-Waterfall-Rebalancing}). 
\item \textbf{capacityToFill} indicates the capacity to fill on the corresponding
asset based on the difference between the current amount on each asset
and maximum capacity on the asset, while taking into account the total
new amount to be deployed or withdrawn. This is explained in Step
(\ref{enu:12-Then-calculate-the}) of Algorithm (\ref{alg:Rebalancing-Algorithm})
in Section (\ref{sec:The-Cascading-Waterfall-Rebalancing}). 
\end{enumerate}
\end{itemize}

\begin{figure}[H]
\includegraphics[width=18cm]{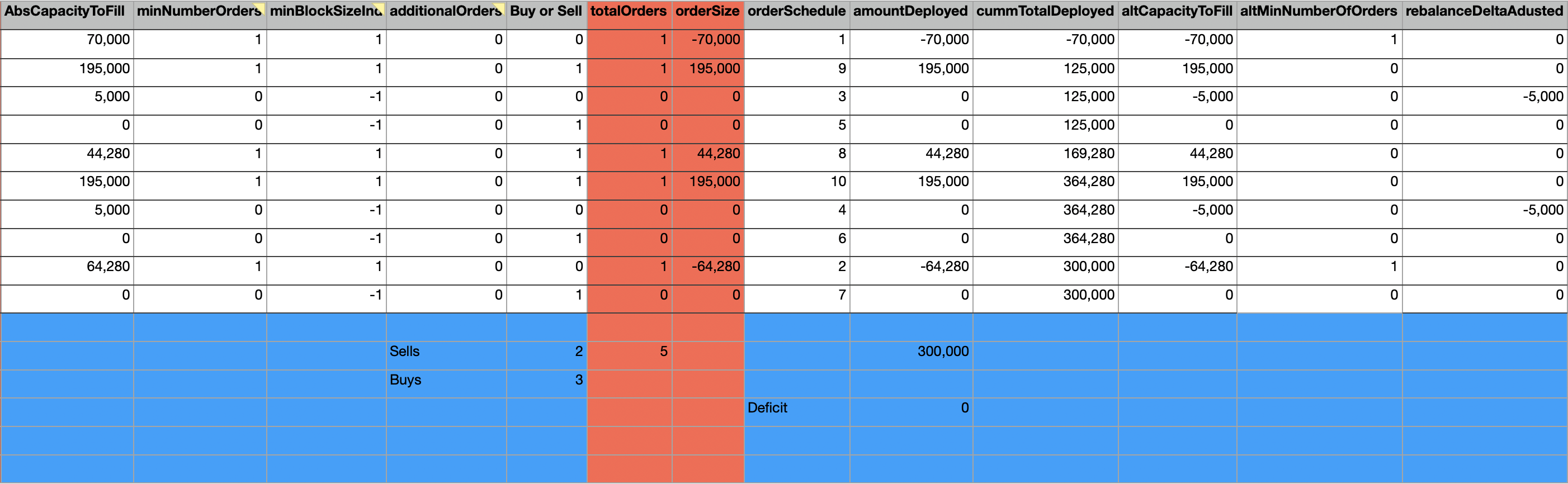}

\caption{\label{fig:Rebalancing-Illustration:-Cascading-Order-Schedule}Rebalancing
Illustration: Cascading Mechanism Order Schedule}
\end{figure}

\begin{itemize}
\item The Table in Figure (\ref{fig:Rebalancing-Illustration:-Cascading-Order-Schedule})
shows numerical examples related to the order schedule generated for
the cascading rebalancing mechanism in Algorithm (\ref{alg:Rebalancing-Algorithm})
based on the material in Section (\ref{sec:The-Cascading-Waterfall-Rebalancing}).
This figure illustrates the number of trades and the size of each
trade across each asset. Sell orders are sent first followed by buy
orders since the amount received from selling assets is required for
investment into assets that require additional inflows.
\item The columns in Figure (\ref{fig:Rebalancing-Illustration:-Cascading-Order-Schedule})
represent the following information respectively: 
\begin{enumerate}
\item \textbf{AbsCapacityToFill} corresponds to each rebalance event corresponding
to 
\item \textbf{minNumberOrders} is the minimum number of orders that need
to be made on this asset based on the cascading rebalancing mechanism
described in Algorithm (\ref{alg:Rebalancing-Algorithm}) in Section
(\ref{sec:The-Cascading-Waterfall-Rebalancing}). This column shows
zero - 0- if the absolute Value of capacity to fill is less than minimum
block size; and 1 otherwise. This is explained in Step (\ref{enu:13-Calculate-the-minimum})
of Algorithm (\ref{alg:Rebalancing-Algorithm}) in Section (\ref{sec:The-Cascading-Waterfall-Rebalancing}). 
\item \textbf{minBlockSizeInd} indicates that the trade size on this asset
is below the minimum trade size for that asset. This column shows
minus one, -1, if the absolute Value of capacity to fill is less than
minimum block size; 1 otherwise. 
\item \textbf{additionalOrders} is the number of additional orders required
to fill the total deficit in this asset. This is the number of orders
in addition to the minimum number of orders indicated by \textbf{minNumberOrders.
}This is explained in Step (\ref{enu:13-Calculate-the-minimum}) of
Algorithm (\ref{alg:Rebalancing-Algorithm}) in Section (\ref{sec:The-Cascading-Waterfall-Rebalancing}). 
\item \textbf{Buy or Sell} indicates if we are making buy or sell trades
on this asset. 
\item \textbf{totalOrders} indicates the total number of orders to be made
on this asset. This is the sum of \textbf{minNumberOrders }and\textbf{
additionalOrders. }This is explained in Step (\ref{enu:13-Calculate-the-minimum})
of Algorithm (\ref{alg:Rebalancing-Algorithm}) in Section (\ref{sec:The-Cascading-Waterfall-Rebalancing}). 
\item \textbf{orderSize} is the size of each trade to be made on this asset.
This is explained in Step (\ref{enu:14-Calculate-the-order}) of Algorithm
(\ref{alg:Rebalancing-Algorithm}) in Section (\ref{sec:The-Cascading-Waterfall-Rebalancing}). 
\item \textbf{orderSchedule} indicates the sequence in which trades for
this asset will have to be executed in comparison to the other assets
in the portfolio.
\item \textbf{amountDeployed} indicates the total amount deployed on this
asset in this rebalancing event. This is the product of \textbf{totalOrders
}and\textbf{ orderSize.}
\item \textbf{cummTotalDeployed} indicates the cumulative total deployed
- summing up the amounts invested or withdrawn, bought or sold - across
all the assets in the portfolio starting from the asset in the first
row.
\item \textbf{altCapacityToFill} is an alternative calculation to find the
capacity to fill - \textbf{capacityToFill }in Figure (\ref{fig:Rebalancing-Illustration:-Cascading-Capacity-Fill})
- on the corresponding asset based on the difference between the current
amount on each asset and maximum capacity on the asset, while taking
into account the total new amount to be deployed or withdrawn. This
is explained in Step (\ref{enu:12-Then-calculate-the}) of Algorithm
(\ref{alg:Rebalancing-Algorithm}) in Section (\ref{sec:The-Cascading-Waterfall-Rebalancing}). 
\item \textbf{altMinNumberOfOrders} is an alternative calculation to find
the minimum number of orders - \textbf{minNumberOrders }in Figure
(\ref{fig:Rebalancing-Illustration:-Cascading-Capacity-Fill}) - but
with the difference that orders are counted only when the quantities
are within the minimum and maximum trade size. That is if the quantity
to be traded on the asset is less than the minimum trade size, it
is not considered as an order.
\item \textbf{rebalanceDeltaAdusted} indicates the rebalance delta - \textbf{rebalanceDelta}
in Figure (\ref{fig:Rebalancing-Illustration:-Cascading-Capacity-Fill})
- on the asset by taking into account the alternate minimum number
of orders, \textbf{altMinNumberOfOrders}.
\end{enumerate}
\end{itemize}

\section{\label{sec:Areas-for-Further}Implementation Pointers and Areas for
Further Research}

The two components - calculation of asset capacities and the rebalancing
methodology described in this paper - are some of the earliest pieces
that need to be built, and tested, to understand how well they apply
to various investment mandates and how seamlessly they would work
in a blockchain environment. Once these initial efforts have proved
satisfactory, attempting to make improvements is quite natural. If
any investment firm has to approach the implementation in terms of
phases, it would be prudent in the initial phase to ensure these techniques
could be invoked and utilized on an on-demand basis. The next set
of enhancements are to be able to connect them to live data updates,
and completely automate them, so that these calculations can run on
a daily basis or even several times during a 24 hour period.

The rebalancing algorithm can be a stand alone component - external
to the blockchain system - that reads the investment variables and
outputs the trade schedule for each asset, which will be used by fund
personnel to perform the necessary trades. The trades need to happen
on the blockchain environment, but the calculations for the steps
in the rebalancing algorithm can be done in off-chain scripts. The
weight calculation engine will be a separate component which interact
with the rebalancing calculation routines and provides the relevant
information to the portfolio teams. A similar stand-alone component
can perform the minimum and maximum trade size calculations. This
ensures that each piece can be enhanced without affecting the other
components.

Numerous alternative weight calculation techniques are possible. We
have used the closing prices to calculate the weights. But open, high
or low prices for a particular day can be used instead of the close
prices to get alternate weights. The rebalancing algorithm looks at
the difference between current amounts invested in an asset from the
minimum and maximum capacities, instead we how the ideal capacities
vary from the current amounts - based on the ideal weights of the
asset - to get the trades to be done on any asset. Again the choice
of which approach should depend on how often we plan to rebalance
and how closely we wish to stay aligned with the calculated weights
and the extent of forecast errors that persist in the operating environment.

To govern a system with many moving parts, such as the one described
in this paper, several parameters need to be monitored and tweaked
on a regular basis. The portfolio management team will have to observe
these parameters continuously and update them, as necessary, using
specialized internal tools. The bulk of the configurations that decide
how the system will run on a periodic basis are related to asset capacities
and trade executions. In addition, trade executions can be error prone
wherein failures need to be monitored and intelligent customizations
to retry need to be incorporated into the process. Hence trade execution
related parameters and operational procedures will garner significant
focus and a big chunk of time from the investment team. 

Internal tools have to be designed such that the flow of funds happens
automatically, for the most part, with human intervention to complement
the decision making. Significant automation of the investment apparatus
will allow the fund to take advantage of market opportunities seamlessly
and human oversight will enable the team to watch out for exceptional
situations and fine tune the decisions. This coupling of man and machine
will lead to a better final outcome for all our participants. 

An illustration of this pairing is that the approach we have described
to investing on blockchain will benefit from volatility, which is
seen as the bane of crypto markets by most players. Volatility, which
is the up and down movement of asset prices, will cause our rebalancing
algorithm to buy assets that drop in prices and sell assets as they
start soaring again. But to filter out the noise and react only to
real signals, the buying and selling happens only when certain boundaries
or range thresholds are crossed. This spectrum over which transactions
happen are automatically calculated based on asset properties, but
have to be fine tuned by investment specialists. Suffice it to say,
while mathematical optimization techniques offer powerful venues to
garner profits, they might fall short of conquering the extreme scenarios
that markets present. Hence mixing mathematics models with human intuition,
that takes care of exceptional cases, is the ideal recipe for wealth
creation.

Our rebalancing methodology will significantly outperform any simple
rebalancing mechanism (without minimum and maximum weights on assets)
over time in terms of gas fees savings and market impact costs. It
could under-perform the simple mechanism when new assets are being
added and removed frequently. The underperformance is small compared
to the benefits. The algorithm we have designed will provide better
performance when the number of assets increases. The algorithm performance
will converge to any simple rebalancing mechanism when the minimum
and maximum weights on the assets get closer to one another. Quantitative
bounds and other analytical results will be provided in a later paper
since this paper has a lot of innovations and material already. But
if necessary, we can include some of the mathematical estimates as
propositions - with the proofs provided in the appendix of this paper
or in the related paper. 

For the sake of brevity, we have focused on the central elements of
our technique. The actual technical implementation will have to cover
several specialized scenarios, nuances or other constraints. Additional
checks pertaining to division by zero and other such cases need to
be considered in the software coding and testing (End-note \ref{enu:Software-Testing-Validation}). 

\section{\label{sec:Conclusion}Conclusions}

We have designed a rebalancing algorithm that is tailored for the
nuances of blockchain based trading and portfolio management. Our
rebalancing mechanism recommends an ideal size and number of trades
for each asset, factoring in the gas fee and slippage. The essence
of the model we have created gives indications regarding whether trades
should be made on individual assets depending on the uncertainty in
the micro - asset level characteristics - and macro - aggregate market
factors - environments. 

The algorithm is presented as a sequence of steps with detailed explanations
for the computations that happen in each step and the rationale for
performing those calculations. We have provided several numerical
examples to illustrate the various steps - including the calculation
of numerous intermediate variables - of our rebalancing approach.
The Algorithm we have developed can be easily applied outside blockchain
to investment funds across all asset classes at any trading frequency
and rebalancing duration. In the hyper-volatile crypto market, our
approach to daily rebalancing will benefit from volatility. Price
movements will cause our algorithm to buy assets that drop in prices
and sell as they soar. In fact, the buying and selling happen only
when certain boundaries are crossed in order to weed out any market
noise and ensure sound trade execution. 

We have also shown different ways to calculate portfolio weights and
the minimum and maximum trade sizes on an asset that depend on the
blockchain network and the properties of the asset. The asset weights
and minimum and maximum trade sizes act as inputs to the rebalancing
algorithm which then outputs the number of trades on each asset and
the size of each trade. We have given detailed mathematical formulations,
and technical pointers, to be able to implement the algorithm and
its related components as stand-alone technical pieces. Each piece
receives information from the blockchain decentralized ledger - and
other independent external data sources - to perform its designated
calculations. The result of the calculations aid the investment strategy
in maintaining its preferred risk and return objectives. Careful orchestration
among mathematical optimization for portfolio construction, trade
automation of the investment apparatus, and human oversight will allow
us to watch out for exceptional situations and ultimately lead to
better risk management - which is very much the need of the hour for
the decentralized financial landscape. 

\section{\label{sec:Explanations-and-End-notes}Explanations and End-notes}
\begin{enumerate}
\item \label{EN:To-Be-Not-To-Be}To be, or not to be\textquotedbl{} is the
opening phrase of a soliloquy given by Prince Hamlet in the so-called
\textquotedbl nunnery scene\textquotedbl{} of William Shakespeare's
play Hamlet, Act 3, Scene 1. (William Shakespeare: \href{https://en.wikipedia.org/wiki/William_Shakespeare}{William Shakespeare, Wikipedia Link})

To be, or not to be, that is the question: 

Whether 'tis nobler in the mind to suffer 

The slings and arrows of outrageous fortune, 

Or to take Arms against a Sea of troubles, 

And by opposing end them: to die, to sleep ...
\item \label{enu:In-finance-Rebalancing}In finance and investing, rebalancing
of investments (or constant mix) is a strategy of bringing a portfolio
that has deviated away from one's target asset allocation back into
line. \href{https://en.wikipedia.org/wiki/Rebalancing_investments}{Rebalancing Investments,  Wikipedia Link}
\item \label{enu:In-finance,-risk-Factors}In finance, risk factors are
the building blocks of investing, that help explain the systematic
returns in equity market, and the possibility of losing money in investments
or business adventures. \href{https://en.wikipedia.org/wiki/Risk_factor_(finance)}{Risk Factor (finance),  Wikipedia Link}
\item \label{enu:Decentralized-finance}Decentralized finance (often stylized
as DeFi) offers financial instruments without relying on intermediaries
such as brokerages, exchanges, or banks by using smart contracts on
a blockchain. \href{https://en.wikipedia.org/wiki/Decentralized_finance}{Decentralized Finance (DeFi), Wikipedia Link}
\item \label{enu:CoinMarketCap}CoinMarketCap is a leading price-tracking
website for crypto-assets in the cryptocurrency space. Its mission
is to make crypto discoverable and efficient globally by empowering
retail users with unbiased, high quality and accurate information
for drawing their own informed conclusions. It was founded in May
2013 by Brandon Chez. \href{https://coinmarketcap.com/about/}{CoinMarketCap, Website Link}
\begin{enumerate}
\item \label{enu:Crypto-Ranking}A ranking of cryptocurrencies, including
symbols for the various tokens, by market capitalization is available
on the CoinMarketCap website. We are using the data as of May-25-2022,
when the first version of this article was written. \href{https://coinmarketcap.com}{CoinMarketCap Cryptocurrency Ranking,  Website Link}
\end{enumerate}
\item \label{enu:Types-Yield-Enhancement-Services}The following are the
four main types of blockchain decentralized financial products or
services. These are sometimes collectively referred to as vaults since
it requires additional steps to withdraw funds from them. We can also
consider them as the main types of yield enhancement, or return generation,
vehicles available in decentralized finance: 
\begin{enumerate}
\item \label{enu:Single-sided-staking-allows}Single-Sided Staking: This
allows users to earn yield by providing liquidity for one type of
asset, in contrast to liquidity provisioning on AMMs, which requires
a pair of assets. \href{https://docs.saucerswap.finance/features/single-sided-staking}{Single Sided Staking,  SuacerSwap Link}
\begin{enumerate}
\item Bancor is an example of a provider who supports single sided staking.
Bancor natively supports Single-Sided Liquidity Provision of tokens
in a liquidity pool. This is one of the main benefits to liquidity
providers that distinguishes Bancor from other DeFi staking protocols.
Typical AMM liquidity pools require a liquidity provider to provide
two assets. Meaning, if you wish to deposit \textquotedbl TKN1\textquotedbl{}
into a pool, you would be forced to sell 50\% of that token and trade
it for \textquotedbl TKN2\textquotedbl . When providing liquidity,
your deposit is composed of both TKN1 and TKN2 in the pool. Bancor
Single-Side Staking changes this and enables liquidity providers to:
Provide only the token they hold (TKN1 from the example above) Collect
liquidity providers fees in TKN1. \href{https://docs.bancor.network/about-bancor-network/faqs/single-side-liquidity}{Single Sided Staking,  Bancor Link}
\end{enumerate}
\item \label{enu:AMM-Liquidity-Pairs}AMM Liquidity Pairs (AMM LP): A constant-function
market maker (CFMM) is a market maker with the property that that
the amount of any asset held in its inventory is completely described
by a well-defined function of the amounts of the other assets in its
inventory (Hanson 2007). \href{https://en.wikipedia.org/wiki/Constant_function_market_maker}{Constant Function Market Maker,  Wikipedia Link}

This is the most common type of market maker liquidity pool. Other
types of market makers are discussed in Mohan (2022). All of them
can be grouped under the category Automated Market Makers. Hence the
name AMM Liquidity Pairs. A more general discussion of AMMs, without
being restricted only to the blockchain environment, is given in (Slamka,
Skiera \& Spann 2012).
\item \label{enu:LP-Token-Staking:}LP Token Staking: LP staking is a valuable
way to incentivize token holders to provide liquidity. When a token
holder provides liquidity as mentioned earlier in Point (\ref{enu:AMM-Liquidity-Pairs})
they receive LP tokens. LP staking allows the liquidity providers
to stake their LP tokens and receive project tokens tokens as rewards.
This mitigates the risk of impermanent loss and compensates for the
loss. \href{https://defactor.com/liquidity-provider-staking-introduction-guide/}{Liquidity Provider Staking,  DeFactor Link}
\begin{enumerate}
\item Note that this is also a type of single sided staking discussed in
Point (\ref{enu:Single-sided-staking-allows}). The key point to remember
is that the LP Tokens can be considered as receipts for the crypto
assets deposits in an AMM LP Point (\ref{enu:AMM-Liquidity-Pairs}).
These LP Token receipts can be further staked to generate additional
yield.
\end{enumerate}
\item \label{enu:Lending:-Crypto-lending}Lending: Crypto lending is the
process of depositing cryptocurrency that is lent out to borrowers
in return for regular interest payments. Payments are typically made
in the form of the cryptocurrency that is deposited and can be compounded
on a daily, weekly, or monthly basis. \href{https://www.investopedia.com/crypto-lending-5443191}{Crypto Lending,  Investopedia Link};
\href{https://defiprime.com/decentralized-lending}{DeFi Lending,  DeFiPrime Link};
\href{https://crypto.com/price/categories/lending}{Top Lending Coins by Market Capitalization,  Crypto.com Link}.
\begin{enumerate}
\item Crypto lending is very common on decentralized finance projects and
also in centralized exchanges. Centralized cryptocurrency exchanges
are online platforms used to buy and sell cryptocurrencies. They are
the most common means that investors use to buy and sell cryptocurrency
holdings. \href{https://www.investopedia.com/tech/what-are-centralized-cryptocurrency-exchanges/}{Centralized Cryptocurrency Exchanges,  Investopedia Link}
\item Lending is a very active area of research both on blockchain and off
chain (traditional finance) as well (Cai 2018; Zeng et al., 2019;
Bartoletti, Chiang \& Lafuente 2021; Gonzalez 2020; Hassija et al.,
2020; Patel et al. , 2020). 
\item Lending is also a highly profitable business in the traditional financial
world (Kashyap 2022). Investment funds, especially hedge funds, engage
in borrowing securities to put on short positions depending on their
investment strategies.Long only investment funds typically supply
securities or lend their assets for a fee.
\item In finance, a long position in a financial instrument means the holder
of the position owns a positive amount of the instrument. \href{https://en.wikipedia.org/wiki/Long_(finance)}{Long Position in Finance,  Wikipedia Link}
\item In finance, being short in an asset means investing in such a way
that the investor will profit if the value of the asset falls. This
is the opposite of a more conventional \textquotedbl long\textquotedbl{}
position, where the investor will profit if the value of the asset
rises. \href{https://en.wikipedia.org/wiki/Short_(finance)}{Short Position in Finance,  Wikipedia Link}
\end{enumerate}
\end{enumerate}
\item \label{enu:A-sling-is}A sling is a projectile weapon typically used
to throw a blunt projectile such as a stone, clay, or lead \textquotedbl sling-bullet\textquotedbl .
It is also known as the shepherd's sling. \href{https://en.wikipedia.org/wiki/Sling_(weapon)}{Sling (Weapon), Wikipedia Link}
\item \label{enu:The-range-based}The range based models we have outlined
for the weights are based on a wider set of techniques termed: Randoptimization.
The limitations of optimization methodologies and the need for range
based methodologies - which introduce randomness in the decision process
- are discussed in detail in the series: ``Fighting Uncertainty with
Uncertainty'' Kabeer (2016). The minimum and maximum asset weights
we have discussed in the main text are based on this idea of operating
a system within a range as opposed to pinning down operational parameters
to a single value. The range of values is prudent to use due to the
errors that exist around the estimates we obtain for an ideal value.
Clearly, the weight range we can use to minimize rebalancing requirements
is dependent upon the estimation errors in the corresponding weight
optimization process. Conversely, depending on the extent to which
we wish to undertake rebalancing - that is how often and the size
of trades in comparison to the holdings in the portfolio - we can
decide the width of the range we can tolerate for the weights.
\item \label{enu:Stablecoin}A Stablecoin is a type of cryptocurrency where
the value of the digital asset is supposed to be pegged to a reference
asset, which is either fiat money, exchange-traded commodities (such
as precious metals or industrial metals), or another cryptocurrency.
\href{https://en.wikipedia.org/wiki/Stablecoin}{Stable Coin,  Wikipedia Link}
\item \label{enu:Software-Testing-Validation}We would like to highlight
the following points to help with the actual coding of the software
(Boehm 1983; Balci 1995; Desikan \& Ramesh 2006; Green \& Ledgard
2011; Knuth 2014). The algorithm we have provided acts mostly as detailed
implementation guidelines. Many cases and error conditions need to
be handled appropriately during implementation. Alternate implementation
simplifications, time conventions, and counters are possible and can
be accommodated accordingly. There might even be some issues - or
bugs - with the variables, counters and timing. These are due to limitations
of not actually testing scenarios using a full fledged software system.
But the gist of what we have provided should carry over to the coding
stage with very little changes. Conditional statements such as - if
... then ... else - can be used depending on the implementation language
and other efficiency considerations as necessary.
\end{enumerate}

\section{\label{sec:References}References}
\begin{itemize}
\item Almgren, R., \& Chriss, N. (2001). Optimal execution of portfolio
transactions. Journal of Risk, 3, 5-40.
\item An, Z., Chen, C., Li, D., \& Yin, C. (2021). Foreign institutional
ownership and the speed of leverage adjustment: International evidence.
Journal of Corporate Finance, 68, 101966.
\item Ante, L., Fiedler, I., \& Strehle, E. (2021). The influence of stablecoin
issuances on cryptocurrency markets. Finance Research Letters, 41,
101867.
\item Balci, O. (1995, December). Principles and techniques of simulation
validation, verification, and testing. In Proceedings of the 27th
conference on Winter simulation (pp. 147-154). 
\item Bartoletti, M., Chiang, J. H. Y., \& Lafuente, A. L. (2021). SoK:
lending pools in decentralized finance. In Financial Cryptography
and Data Security. FC 2021 International Workshops: CoDecFin, DeFi,
VOTING, and WTSC, Virtual Event, March 5, 2021, Revised Selected Papers
25 (pp. 553-578). Springer Berlin Heidelberg. 
\item Bertsimas, D., \& Lo, A. W. (1998). Optimal control of execution costs.
Journal of financial markets, 1(1), 1-50.
\item Boehm, B. W. (1983). Seven basic principles of software engineering.
Journal of Systems and Software, 3(1), 3-24. 
\item Bouchey, P., Nemtchinov, V., Paulsen, A., \& Stein, D. M. (2012).
Volatility harvesting: Why does diversifying and rebalancing create
portfolio growth?. The Journal of Wealth Management, 15(2), 26-35. 
\item Bradley, A.C. (1991). Shakespearean Tragedy: Lectures on Hamlet, Othello,
King Lear and Macbeth. London: Penguin. ISBN 978-0-14-053019-3.
\item Cai, C. W. (2018). Disruption of financial intermediation by FinTech:
a review on crowdfunding and blockchain. Accounting \& Finance, 58(4),
965-992. 
\item Calvet, L. E., Campbell, J. Y., \& Sodini, P. (2009). Fight or flight?
Portfolio rebalancing by individual investors. The Quarterly journal
of economics, 124(1), 301-348.
\item Camanho, N., Hau, H., \& Rey, H. (2017). Global portfolio rebalancing
under the microscope. Review of Financial Studies.
\item Camanho, N., Hau, H., \& Rey, H. (2022). Global portfolio rebalancing
and exchange rates. The Review of Financial Studies, 35(11), 5228-5274. 
\item Cernera, F., La Morgia, M., Mei, A., \& Sassi, F. (2023). Token Spammers,
Rug Pulls, and Sniper Bots: An Analysis of the Ecosystem of Tokens
in Ethereum and in the Binance Smart Chain (\{\{\{\{\{BNB\}\}\}\}\}).
In 32nd USENIX Security Symposium (USENIX Security 23) (pp. 3349-3366). 
\item Chambers, D. R., \& Zdanowicz, J. S. (2014). The limitations of diversification
return. The Journal of Portfolio Management, 40(4), 65-76.
\item Chauhan, G. S., \& Huseynov, F. (2018). Corporate financing and target
behavior: New tests and evidence. Journal of Corporate Finance, 48,
840-856.
\item Cochrane, J. (2009). Asset pricing: Revised edition. Princeton university
press.
\item Cook, D. O., \& Tang, T. (2010). Macroeconomic conditions and capital
structure adjustment speed. Journal of corporate finance, 16(1), 73-87.
\item Cuthbertson, K., Hayley, S., Motson, N., \& Nitzsche, D. (2016). What
does rebalancing really achieve?. International Journal of Finance
\& Economics, 21(3), 224-240. 
\item Desikan, S., \& Ramesh, G. (2006). Software testing: principles and
practice. Pearson Education India. 
\item Donmez, A., \& Karaivanov, A. (2022). Transaction fee economics in
the Ethereum blockchain. Economic Inquiry, 60(1), 265-292.
\item Elton, E. J., Gruber, M. J., Brown, S. J., \& Goetzmann, W. N. (2009).
Modern portfolio theory and investment analysis. John Wiley \& Sons.
\item Gonzalez, L. (2020). Blockchain, herding and trust in peer-to-peer
lending. Managerial Finance, 46(6), 815-831. 
\item Green, R., \& Ledgard, H. (2011). Coding guidelines: Finding the art
in the science. Communications of the ACM, 54(12), 57-63. 
\item Guastaroba, G., Mansini, R., \& Speranza, M. G. (2009). Models and
simulations for portfolio rebalancing. Computational Economics, 33,
237-262.
\item Hallerbach, W. G. (2014). Disentangling rebalancing return. Journal
of Asset Management, 15(5), 301-316.
\item Hanson, R. (2007). Logarithmic market scoring rules for modular combinatorial
information aggregation. The Journal of Prediction Markets, 1(1),
3-15. 
\item Hassija, V., Bansal, G., Chamola, V., Kumar, N., \& Guizani, M. (2020).
Secure lending: Blockchain and prospect theory-based decentralized
credit scoring model. IEEE Transactions on Network Science and Engineering,
7(4), 2566-2575. 
\item Jensen, J. R., von Wachter, V., \& Ross, O. (2021). An introduction
to decentralized finance (defi). Complex Systems Informatics and Modeling
Quarterly, (26), 46-54.
\item Juelsrud, R. E., \& Wold, E. G. (2020). Risk-weighted capital requirements
and portfolio rebalancing. Journal of Financial Intermediation, 41,
100806.
\item Kabeer, B. (2016). Fighting Uncertainty with Uncertainty. Available
at SSRN 2715424.
\item Karastilo, P. (2020). David vs Goliath (You against the Markets),
A dynamic programming approach to separate the impact and timing of
trading costs. Physica A: Statistical Mechanics and its Applications,
545, 122848.
\item Kashyap, R. (2022). Bringing Risk Parity To The DeFi Party: A Complete
Solution To The Crypto Asset Management Conundrum. Working Paper.
\item Kashyap, R. (2023). Arguably Adequate Aqueduct Algorithm: Crossing
A Bridge-Less Block-Chain Chasm. Finance Research Letters, 58, 104421.
\item Katarina, A. (2023). DeFi Security: Turning The Weakest Link Into
The Strongest Attraction. arXiv preprint arXiv:2312.00033.
\item Kimball, M. S., Shapiro, M. D., Shumway, T., \& Zhang, J. (2020).
Portfolio rebalancing in general equilibrium. Journal of Financial
Economics, 135(3), 816-834.
\item Knuth, D. E. (2014). Art of computer programming, volume 2: Seminumerical
algorithms. Addison-Wesley Professional.
\item Leary, M. T., \& Roberts, M. R. (2005). Do firms rebalance their capital
structures?. The journal of finance, 60(6), 2575-2619.
\item Maeso, J. M., \& Martellini, L. (2020). Measuring portfolio rebalancing
benefits in equity markets. The Journal of Portfolio Management, 46(4),
94-109. 
\item Masters, S. J. (2003). Rebalancing. The Journal of Portfolio Management,
29(3), 52-57. 
\item Mohan, V. (2022). Automated market makers and decentralized exchanges:
a DeFi primer. Financial Innovation, 8(1), 20.
\item Patel, S. B., Bhattacharya, P., Tanwar, S., \& Kumar, N. (2020). Kirti:
A blockchain-based credit recommender system for financial institutions.
IEEE Transactions on Network Science and Engineering, 8(2), 1044-1054. 
\item Pierro, G. A., Rocha, H., Tonelli, R., \& Ducasse, S. (2020, February).
Are the gas prices oracle reliable? a case study using the ethgasstation.
In 2020 IEEE International Workshop on Blockchain Oriented Software
Engineering (IWBOSE) (pp. 1-8). IEEE.
\item Pierro, G. A., Rocha, H., Ducasse, S., Marchesi, M., \& Tonelli, R.
(2022). A user-oriented model for oracles’ gas price prediction. Future
Generation Computer Systems, 128, 142-157.
\item Qian, E. (2012). Diversification return and leveraged portfolios.
The Journal of Portfolio Management, 38(4), 14-25.
\item Rastad, M. (2016). Capital structure pre-balancing: Evidence from
convertible bonds. Journal of Corporate Finance, 41, 43-65.
\item Rattray, S., Granger, N., Harvey, C. R., \& Van Hemert, O. (2020).
Strategic rebalancing. The Journal of Portfolio Management, 46(6),
10-31.
\item Ross, S. A., Westerfield, R., \& Jaffe, J. F. (1999). Corporate finance.
Irwin/McGraw-Hill.
\item Slamka, C., Skiera, B., \& Spann, M. (2012). Prediction market performance
and market liquidity: A comparison of automated market makers. IEEE
Transactions on Engineering Management, 60(1), 169-185. 
\item Stein, D. M., Nemtchinov, V., \& Pittman, S. (2009). Diversifying
and rebalancing emerging market countries. The Journal of Wealth Management,
11(4), 79.
\item Sun, W., Fan, A., Chen, L., Schouwenaars, T., \& Albota, M. A. (2006).
Optimal rebalancing for institutional portfolios. Journal of Portfolio
Management, 32(2), 33.
\item Stutzer, M. (2010). The paradox of diversification. The Journal of
Investing, 19(1), 32-35.
\item Tokat, Y., \& Wicas, N. W. (2007). Portfolio rebalancing in theory
and practice. The Journal of Investing, 16(2), 52-59. 
\item Willenbrock, S. (2011). Diversification return, portfolio rebalancing,
and the commodity return puzzle. Financial Analysts Journal, 67(4),
42-49. 
\item Woodside-Oriakhi, M., Lucas, C., \& Beasley, J. E. (2013). Portfolio
rebalancing with an investment horizon and transaction costs. Omega,
41(2), 406-420. 
\item Xu, J., \& Feng, Y. (2022). Reap the Harvest on Blockchain: A Survey
of Yield Farming Protocols. IEEE Transactions on Network and Service
Management.
\item Yu, J. R., \& Lee, W. Y. (2011). Portfolio rebalancing model using
multiple criteria. European Journal of Operational Research, 209(2),
166-175.
\item Zarir, A. A., Oliva, G. A., Jiang, Z. M., \& Hassan, A. E. (2021).
Developing cost-effective blockchain-powered applications: A case
study of the gas usage of smart contract transactions in the ethereum
blockchain platform. ACM Transactions on Software Engineering and
Methodology (TOSEM), 30(3), 1-38.
\item Zeng, X., Hao, N., Zheng, J., \& Xu, X. (2019). A consortium blockchain
paradigm on hyperledger-based peer-to-peer lending system. China Communications,
16(8), 38-50.
\item Zetzsche, D. A., Arner, D. W., \& Buckley, R. P. (2020). Decentralized
finance (defi). Journal of Financial Regulation, 6, 172-203.
\end{itemize}

\end{document}